\documentclass[12pt,a4paper,english,pointlessnumbers]{scrartcl}
\usepackage[T1]{fontenc}
\usepackage[latin1]{inputenc}
\usepackage{graphicx}
\usepackage{amssymb}

\makeatletter

\providecommand{\tabularnewline}{\\}

\usepackage{amsmath,amssymb,amsfonts,amscd,bbm,mathrsfs}
\usepackage{graphicx}
\usepackage{cite}
\usepackage{lastpage}

\usepackage{babel}
\makeatother
\begin{document}
\thispagestyle{empty} \renewcommand{\thefootnote}{\fnsymbol{footnote}} \setcounter{footnote}{1}

\vspace*{-1.cm}

\begin{flushright}TUM-HEP-527/03, OSU-HEP-03-12 \end{flushright}

\vspace*{1.8cm}

\begin{center}\textsf{\textbf{\Large Casimir Energy in Deconstruction
}}\\
{\Large  \vspace*{3mm}}\textsf{\textbf{\Large and the Cosmological
Constant}}\end{center}{\Large \par}

\vspace*{14mm}

\centerline{Florian Bauer{\large }%
\footnote{E-mail: \texttt{fbauer@ph.tum.de}%
}, Manfred Lindner%
\footnote{E-mail: \texttt{lindner@ph.tum.de}%
}}

%\vspace*{4mm}

\begin{center}\emph{Institut für Theoretische Physik, Physik-Department,}\\
 \emph{Technische Universität München,}\\
 \emph{James-Franck-Stra\ss{}e, 85748 Garching, Germany}\end{center}

\vspace*{4mm}

\begin{center}Gerhart Seidl%
\footnote{E-mail: \texttt{gseidl@hep.phy.okstate.edu}%
}\end{center}

%\vspace*{4mm}

\begin{center}\emph{Department of Physics, }\\
\emph{Oklahoma State University, }\\
\emph{Stillwater, OK 74078, USA}\end{center}

\vspace*{20mm}

\centerline{\textsf{\textbf{Abstract}}}

We demonstrate that by employing the correspondence between gauge
theories in geometric and in deconstructed extra dimensions, it is
possible to transfer the methods for calculating finite Casimir energy
densities in higher dimensions to the four-dimensional deconstruction
setup. By this means, one obtains an unambiguous and well-defined
prescription to determine finite vacuum energy contributions of four-dimensional
quantum fields which have a higher-dimensional correspondence. Thereby,
large kink masses lead to an exponentially suppressed Casimir effect.
For a specific model, we hence arrive at a small and positive contribution
to the cosmological constant in agreement with observations.

\renewcommand{\thefootnote}{\arabic{footnote}} \setcounter{footnote}{0}

\newpage

\renewcommand{\textrm}{\text}

%fancy aktivieren

%\lhead{\sf{\nouppercase{\rightmark}}} 

%\chead{} \rhead{} 

%\lhead{}

%\lfoot{\today} \cfoot{} \rfoot{Seite \thepage/\pageref{LastPage}} %\renewcommand{\headrulewidth}{0.0pt}

\newcommand{\kl}[1]{{\textstyle #1}}

\section{Introduction\label{sec:Introduction}}

The idea of Kaluza-Klein~(KK) compactification~\cite{Kaluza:1921tu}
of extra spatial dimensions offers the attractive possibility to obtain
realistic four-dimensional~(4D) gauge theories from a simpler higher-dimensional
setup~\cite{Cremmer:1978ds}. In this approach, the 4D theory which
emerges after dimensional reduction is generally characterized by
a tower of KK modes \cite{Antoniadis:1990ew}. Here, the maximum number
of KK modes is restricted by an ultraviolet~(UV) cutoff which reflects
the fact that non-Abelian gauge theories in higher dimensions are
non-renormalizable. Although this leads below the cutoff to a renormalizable
effective 4D theory, the full higher-dimensional gauge-invariance
is in general lost. Recently, however, a class of manifestly gauge-invariant
and renormalizable 4D gauge theories has been proposed~\cite{Arkani-Hamed:2001ca,Hill:2000mu}
which reproduce higher-dimensional physics in their infrared~(IR)
limit. These {}``deconstructed'' higher-dimensional gauge theories%
\footnote{For an early approach in the context of infinite arrays of gauge theories
see, \emph{e.g.} \cite{Halpern:1975yj}%
} use the transverse lattice technique~\cite{Bardeen:1976tm} as a
gauge-invariant regulator to describe the extra dimensions and can
be viewed as viable UV completions of theories in more than four space-time
dimensions~\cite{Arkani-Hamed:2001nc}.

One interesting aspect of compactified extra dimensions is, that quantum
fields in such a non-trivial space-time give rise to the Casimir effect~\cite{Casimir:dh},
inducing a non-vanishing finite vacuum energy. The associated Casimir
force can be attractive and contract the compactified extra dimensions
to a size which is sufficiently small so as to have escaped experimental
detection so far~\cite{Appelquist:1982zs}. Upon integrating out
the extra dimensions, the Casimir energies have additionally the interesting
property that they appear as an effective cosmological constant~$\Lambda$
or \emph{vacuum energy} in the 4D subspace. Indeed, recent cosmological
and astrophysical observations~\cite{Perlmutter:1998np} indicate
that the universe is currently in an accelerating expansion phase
which is most likely driven by a positive~$\Lambda$. To be near
to the observational value~$\rho_{\textrm{obs}}\sim10^{-47}\,\textrm{GeV}^{4}$
of the vacuum energy density in the universe, the generation of~$\Lambda$
via Casimir energies requires a compactification radius~$R/(2\pi)$
in the sub-mm range~\cite{Milton:2002hx}. Actually, in the model
of Arkani-Hamed, Dimopoulos, and Dvali~(ADD) for compactified extra
dimensions~\cite{Arkani-Hamed:1998rs}, the fundamental scale~$M_{*}$
of quantum gravity may be lowered from the 4D Planck scale~$M_{\textrm{Pl}}\sim10^{19}\,\textrm{GeV}$
down to the TeV scale, when the compactification radius is of sub-mm
size. Such {}``large'' extra dimensions may be possible if all Standard
Model~(SM) gauge and other degrees of freedom are confined to a 4D
subspace, \emph{i.e.}, on a {}``3-brane'', which can be understood
in certain types of string theory~\cite{Antoniadis:1998ig}. Additional
SM singlet fields, on the other hand, can freely propagate in the
bulk and give rise to a characteristic mixing pattern with SM fields~\cite{Arkani-Hamed:1998vp}.
For gravity freely propagating in the bulk, however, large extra dimensions
are already close to be ruled out by tests of the theory of gravity~\cite{Adelberger:2003zx}
and supernova emission of KK-gravitons~\cite{Cullen:1999hc} which
in total implies that the compactification radius is smaller than~$100\,\mu\textrm{m}$.
In most formulations of deconstruction, on the other hand, gravity
is completely decoupled. A priori, a deconstructed version of the
ADD-scheme can therefore give a realistic value of~$\Lambda$ via
the Casimir effect without running into conflict with the bounds from
gravitational physics.

In a flat Minkowski world, quantum fields also provide a vacuum energy
in the form of infinite zero-point energies. Unfortunately, in absence
of a characteristic length scale like, \emph{e.g.}, the compactification
scale in KK theories, a procedure to calculate a reasonable value
of this energy is not known. Naive estimations using common particle
physics scales as cutoff yield only unrealistically high values for
the vacuum energy density which constitutes one part of the cosmological
constant problem~\cite{Weinberg:1988cp}. Since the effective Lagrangians
of KK modes provided by deconstruction are also defined in Minkowski
space-time, one could, at first instance, expect these theories to
suffer from similar problems of zero-point energies. However, in this
paper we demonstrate that by insisting on the correspondence between
gauge theories in geometric and in deconstructed extra dimensions,
it is possible to transfer the methods for calculating finite Casimir
energy densities in higher dimensions to the 4D deconstruction setup.
By this means, one obtains an unambiguous and well-defined prescription
to determine finite vacuum energies of 4D quantum fields which have
a higher-dimensional correspondence. Here, we propose that the smallness
of~$\Lambda$ can be achieved by a replicated type-II seesaw mechanism~\cite{Mohapatra:1980ia,Ma:1998dx}
which {}``naturally'' generates the large length scale~$R\sim100\,\mu\textrm{m}$
in the deconstructed theory. This has the advantage, that we need
only a small number of KK modes to obtain a realistic~$\Lambda$,
which is in contrast to a naive latticization of the ADD-scheme, where
a short distance cutoff of order~$a^{-1}\sim1\,\textrm{TeV}$ would
require a rather large number of~$N=Ra\sim10^{12}$ lattice sites.
In particular, we calculate the Casimir energies for fields of different
spin, which propagate in a compactified latticized 5th dimension.
As a result, we find that a bulk-Dirac particle gives a value for~$\Lambda$
in agreement with observations. In addition, we show that unwanted
contributions to~$\Lambda$ from quantum fields with large kink (or
bulk) masses are exponentially suppressed and can hence be neglected. 

The paper is organized as follows: First, in Sec.~\ref{sec:Model},
we formulate the model for deconstructed large extra dimensions. Next,
we calculate in Sec.~\ref{sec:ZeroPointCC} the Casimir energies
of scalar~(Sec.~\ref{sub:CasimirScalar}) and fermionic~(Sec.~\ref{sub:CasimirFermion})
bulk fields which are propagating in a latticized 5th dimension. Then,
we analyze in Sec.~\ref{sub:OffsetMass} the exponential suppression
of the unwanted Casimir energies by a kink mass. These results are
related to the model of dimensional deconstruction in Sec.~\ref{sec:ZeroPointDeconst}.
In Sec.~\ref{sec:CasimirAnalytic}, we match the lattice-calculations
onto the continuum theory. Finally, in Sec.~\ref{sec:SummaryConclusions},
we present our summary and conclusions. Additionally, we minimize
in Appendix~\ref{secA:MinimizPotential} the scalar potential involved
in deconstruction. Moreover, explicit calculations for the energy
density, pressure, and the renormalization of these quantities are
presented in Appendix~\ref{secA:ScalarEngPress} and Appendix~\ref{secA:Renormalization}.

%\newpage

\section{Deconstructing Large Extra Dimensions\label{sec:Model}}

In this section, we present first the model for deconstructed large
extra dimensions. Then, we discuss the 5D kinetic terms for gauge
bosons, fermions, and scalars before we outline some phenomenological
features of embedding the deconstructed space into a disk.

\minisec{Sub-mm lattice spacings}

\begin{figure}
\begin{center}\includegraphics[%
  bb=118bp 535bp 382bp 673bp,
  clip]{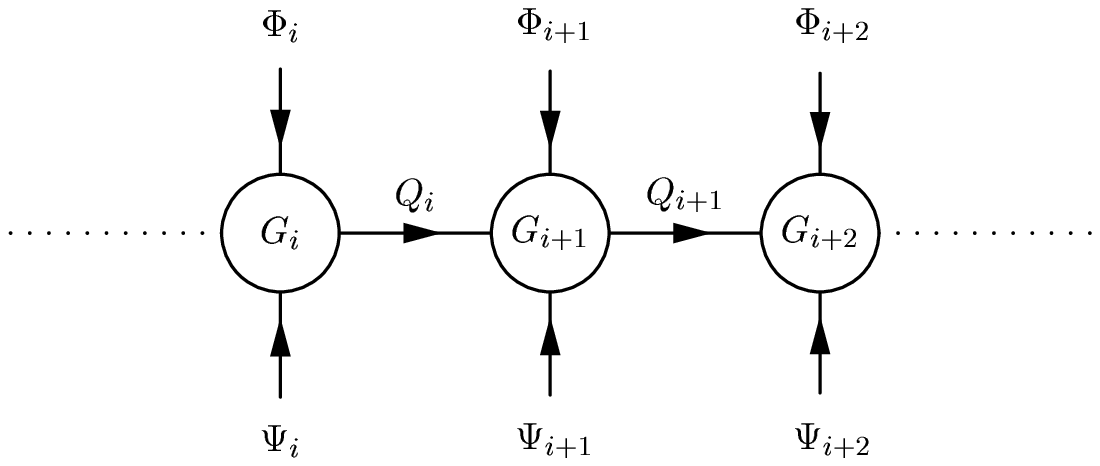}\end{center}

\caption{\label{fig:typeII}View at the sites~$i,i+1$, and~$i+2$ of the
moose diagram for a deconstructed large extra dimension compactified
on~$\mathcal{S}^{1}$. Each circle corresponds to one $U(1)_{i}\equiv G_{i}$
gauge group and each arrow pointing towards (outwards) a circle represents
a field with negative (positive) charge under this group.}
\end{figure}

Let us start with the periodic model for a deconstructed 5D $U(1)$
gauge theory compactified on the circle~$\mathcal{S}^{1}$~\cite{Arkani-Hamed:2001ca,Hill:2000mu}.
The setup is defined by an~$U(1)^{N}=\Pi_{i=1}^{N}U(1)_{i}$ product
gauge group with~$N$ scalar link variables~$Q_{i}$ $(i=1,\ldots,N)$,
where the link field~$Q_{i}$ carries the~$U(1)$-charges~$(q,-q)$
under the neighboring groups~$U(1)_{i}\times U(1)_{i+1}$. The identification~$i+N=i$
establishes the periodicity of the lattice.%
\footnote{To account for twisted quantum fields, we will consider in Sec.~\ref{sub:CasimirScalar}
an anti-periodic lattice with the condition~$i+N=-i$.%
} On the $i$th lattice site, we put one Dirac fermion~$\Psi_{i}$
and one scalar~$\Phi_{i}$ which carry both the charge~$-q$ of
the group $U(1)_{i}$. Here, the fermions~$\Psi_{i}$ are SM-singlets
and correspond to a right-handed bulk neutrino in the ADD scheme~\cite{Arkani-Hamed:1998vp}.
The corresponding {}``moose''~\cite{geor86} or {}``quiver''~\cite{douglas:1996xx}
diagram is shown in Fig.~\ref{fig:typeII}. The Lagrangian of this
field theory can be split into several parts,\[
\mathcal{L}=\mathcal{L}_{\textrm{kin}}[\Phi_{i},Q_{i}]+\mathcal{L}_{\textrm{kin}}[A_{i}^{\mu}]+\mathcal{L}_{\textrm{kin}}[\Psi_{i}]+\mathcal{L}_{\textrm{mass}}[\Psi_{i},Q_{i}]-V,\]
where~$\mathcal{L}_{\textrm{kin}}[A_{i}^{\mu}]=\sum_{n=1}^{N}-\frac{1}{4}(\partial_{\mu}A_{n\nu}-\partial_{\nu}A_{n\mu})^{2}$
and~$\mathcal{L}_{\textrm{kin}}[\Psi_{i}]=\sum_{n=1}^{N}\overline{\Psi}_{n}\gamma_{\mu}(\partial^{\mu}-ig_{n}A_{n}^{\mu})\Psi_{n}$
are the standard kinetic terms for the gauge bosons~$A_{i}^{\mu}$
and the fermions~$\Psi_{i}$, respectively. Here,~$\mathcal{L}_{\textrm{kin}}[\Phi_{i},Q_{i}]$
denotes the kinetic terms for the scalars~$\Phi_{i}$ and~$Q_{i}$,
which provide the gauge boson masses. Moreover, we combine the mass
and mixing terms involving the fermions~$\Psi_{i}$ and the link
fields~$Q_{i}$ into~$\mathcal{L}_{\textrm{mass}}[\Psi_{i},Q_{i}]$.

The most general renormalizable scalar potential~$V$ consistent
with the symmetries reads \begin{eqnarray}
V & = & \sum_{i=1}^{N}\left[m^{2}\Phi_{i}^{\dagger}\Phi_{i}+M^{2}Q_{i}^{\dagger}Q_{i}+\frac{1}{2}\lambda_{1}(\Phi_{i}^{\dagger}\Phi_{i})^{2}+\frac{1}{2}\lambda_{2}(Q_{i}^{\dagger}Q_{i})^{2}\right.\nonumber \\
 &  & +\lambda_{3}Q_{i}^{\dagger}Q_{i}\sum_{j=1}^{N}\Phi_{j}^{\dagger}\Phi_{j}+\mu\Phi_{i}Q_{i}\Phi_{i+1}^{\dagger}+\mu^{\ast}\Phi_{i+1}Q_{i}^{\dagger}\Phi_{i}^{\dagger}\nonumber \\
 &  & +\lambda_{4}\Phi_{i}^{\dagger}\Phi_{i}\sum_{j\neq i}\Phi_{j}^{\dagger}\Phi_{j}+\lambda_{5}Q_{i}^{\dagger}Q_{i}\sum_{j\neq i}Q_{j}^{\dagger}Q_{j}\nonumber \\
 &  & +\lambda_{6}Q_{i}Q_{i+1}\Phi_{i}\Phi_{i+2}^{\dagger}+\lambda_{6}^{\ast}Q_{i+1}^{\dagger}Q_{i}^{\dagger}\Phi_{i+2}\Phi_{i}^{\dagger}\big{]},\label{eq:seesawpotential}\end{eqnarray}
 where~$\lambda_{1},\lambda_{2},\ldots,\lambda_{5}$ are dimensionless
real parameters of order unity and~$\lambda_{6}$ is a complex-valued
order unity coefficient. In Eq.~(\ref{eq:seesawpotential}), we can
take the dimensionful quantities~$m$ and~$\mu$ to be of the order
of the electroweak scale~$\left|m\right|,\left|\mu\right|\simeq10^{2}\:{\textrm{GeV}}$
and we take the mass~$M$ of the link fields to be very large, \textit{i.e.},~$\left|M\right|\gg\left|m\right|,\left|\mu\right|$.
Moreover, the square~$m^{2}$ is chosen to be negative while~$M^{2}$
is positive in order to obtain spontaneous symmetry breaking~(SSB).
Note that for a supersymmetric case, the term~$\lambda_{6}$ would
have to vanish at the renormalizable level due to the holomorphy of
the superpotential and the phase of~$\mu$ could be absorbed into
the Yukawa couplings of the fermions~$\Psi_{i}$. In the following,
the parameters~$\mu$ and~$\lambda_{6}$ are therefore chosen to
be real and~$\mu<0$. %
We minimize the potential by going to the real basis \begin{equation}
Q_{i}=q_{i}^{a}+{\textrm{i}}q_{i}^{b}\longrightarrow\left(\begin{matrix}q_{i}^{a}\\
q_{i}^{b}\end{matrix}\right),\quad\Phi_{i}=\phi_{i}^{a}+{\textrm{i}}\phi_{i}^{b}\longrightarrow\left(\begin{matrix}\phi_{i}^{a}\\
\phi_{i}^{b}\end{matrix}\right),\label{eq:realBasis}\end{equation}
where we are interested in a minimum of~$V$ with the following vacuum
structure \begin{equation}
\langle Q_{i}\rangle=\left(\begin{matrix}u\\
0\end{matrix}\right),\quad\langle\Phi_{i}\rangle=\left(\begin{matrix}v\\
0\end{matrix}\right),\quad i=1,2,\ldots,N,\label{eq:universalVEVs}\end{equation}
 \textit{i.e.}, all link variables~$Q_{i}$ have a real universal
vacuum expectation value~(VEV)~$u$ and all site variables~$\Phi_{i}$
have a real universal VEV~$v$. The conditions for an extremum of
the potential~$V$ are~$\partial V/\partial\phi_{i}^{a,b}=\partial V/\partial q_{i}^{a,b}=0$,
where the explicit forms are given in the Eqs.~(\ref{eq:derivVphia}),
(\ref{eq:derivVphib}), (\ref{eq:derivVqa}), and (\ref{eq:derivVqb})
in Appendix~\ref{secA:MinimizPotential}. Requiring~$\partial V/\partial\phi_{i}^{a}=0$
leads for the VEVs in Eq.~(\ref{eq:universalVEVs}) to the minimization
condition \begin{equation}
m^{2}+\left[\lambda_{1}+(N-1)\lambda_{4}\right]v^{2}+(N\lambda_{3}+\frac{1}{2}\lambda_{6})u^{2}+\mu u=0,\label{eq:condition1}\end{equation}
 where~$\langle\partial V/\partial\phi_{i}^{b}\rangle=0$ is automatically
fulfilled for these VEVs. Demanding~$\partial V/\partial q_{i}^{a}=0$
yields with Eq.~(\ref{eq:universalVEVs}) the minimization condition\begin{equation}
u\left[M^{2}+\left(\lambda_{2}+(N-1)\lambda_{5}\right)u^{2}+(N\lambda_{3}+\frac{1}{2}\lambda_{6})v^{2}\right]+\frac{1}{2}\mu v^{2}=0,\label{eq:condition2}\end{equation}
 and~$\langle\partial V/\partial q_{i}^{b}\rangle=0$ is again satisfied
for these VEVs. Solving Eq.~(\ref{eq:condition1}) for $v^{2}$ and
substituting $v^{2}$ into Eq.~(\ref{eq:condition2}), we obtain
a cubic equation for~$u$, which has the real solution\begin{equation}
u=\frac{m^{2}\mu}{2\left[\lambda_{1}+(N-1)\lambda_{4}\right]M^{2}}+\mathcal{O}(M^{-4}),\label{eq:SmallVEVu}\end{equation}
where we have expanded for large~$M$. From Eq.~(\ref{eq:SmallVEVu})
we conclude that for large $M\gg v,m$ and moderate $N$, one obtains
a naturally small and positive value for $u$, since the VEVs of the
link variables are suppressed via the type-II seesaw mechanism%
\footnote{The hierarchy~$m,\mu\ll M$ is part of this mechanism. Therefore
stability against quantum corrections is achieved as long as the seesaw
type-II mechanism is operative.%
} \cite{Mohapatra:1980ia,Ma:1998dx}. Furthermore, from the Eqs.~(\ref{eq:condition1})
and~(\ref{eq:SmallVEVu}), one finds \begin{equation}
v^{2}=\frac{-m^{2}}{\lambda_{1}+(N-1)\lambda_{4}}+\mathcal{O}(M^{-2}).\label{eq:LargeVEVv}\end{equation}
Choosing the masses of the link fields in the range~$M\simeq10^{8}\ldots10^{9}\:{\textrm{GeV}}$,
we therefore obtain~$v\simeq10^{2}\:{\textrm{GeV}}$ and a seesaw
suppressed value~$u\simeq10^{-1}\ldots10^{-3}\:{\textrm{eV}}$ of
the inverse lattice spacing which corresponds to a sub-mm separation
of the lattice sites.

\minisec{Kinetic and mass terms}

The mass spectrum of the gauge bosons~$A_{i\mu}$ arises via the
Higgs mechanism from the kinetic terms of the scalars~$\Phi_{i}$
and~$Q_{i}$: \begin{eqnarray}
\mathcal{L}_{\textrm{kin}}[\Phi_{i},Q_{i}] & = & \sum_{i=1}^{N}\left[(D_{\mu}\Phi_{i})^{\dag}D^{\mu}\Phi_{i}+(D_{\mu}Q_{i})^{\dag}D^{\mu}Q_{i}\right]\nonumber \\
 & = & \sum_{i=1}^{N}\left|(\partial_{\mu}+ig_{i}A_{i\mu})\Phi_{i}\right|^{2}+\left|(\partial_{\mu}+ig_{i}A_{i\mu}-ig_{i+1}A_{(i+1)\mu})Q_{i}\right|^{2}.\label{eq:LkinPhiQ}\end{eqnarray}
Let us now examine in some more detail how~$\mathcal{L}_{\textrm{kin}}[\Phi_{i},Q_{i}]$
reproduces a 5D~$U(1)$ gauge theory compactified on ~$\mathcal{S}^{1}$.
For this purpose, we denote by~$(x^{\mu},y)$ the bulk coordinates
and by~$A_{5}$ the 5th component of the bulk~$U(1)$ gauge group.
When the fields~$Q_{n}$ are interpreted as the Higgs links%
\footnote{For a detailed discussion of deconstructed 5D QED see, \emph{e.g.},
Ref.~\cite{Hill:2002me}.%
}\begin{equation}
Q_{n}=\frac{u}{\sqrt{2}g_{n}}\exp\left(ig_{n}\int_{an}^{a(n+1)}\textrm{d}y\, A_{5}(x^{\mu},y)\right)=\frac{u}{\sqrt{2}g_{n}}\exp(ig_{n}aA_{n5}),\label{eq:QnIntegralDX5}\end{equation}
where~$a$ is the spacing between neighboring branes, we observe
that~$\sum_{i=1}^{N}(D_{\mu}Q_{i})^{\dag}D^{\mu}Q_{i}$ becomes a
lattice approximation of the 5D gauge kinetic term\[
\frac{1}{a}\sum_{i=1}^{N}(D_{\mu}Q_{i})^{\dag}D^{\mu}Q_{i}\longrightarrow-\frac{1}{4}\int_{0}^{R}\textrm{d}y\,(\partial_{\mu}A_{5}-\partial_{5}A_{\mu})^{2},\,\,\,\, R=Na.\]
Actually, in the non-linear sigma model approximation,~$Q_{n}$ can
be written as\begin{equation}
Q_{n}=\frac{u}{\sqrt{2}g_{n}}\exp(ig_{n}\pi_{n}(x^{\mu})/u),\label{eq:QnNonLinSigma}\end{equation}
where~$\pi_{n}(x^{\mu})$ is the Nambu-Goldstone boson field associated
with~$Q_{n}$. Comparison with Eq.~(\ref{eq:QnIntegralDX5}) shows
that the effective physical degrees of freedom of the non-linear sigma
model field in Eq.~(\ref{eq:QnNonLinSigma}) are completely captured
by the gauge boson and (pseudo) Nambu-Goldstone boson sectors. For
universal gauge couplings~$g_{i}=g$, we obtain from the last line
in Eq.~(\ref{eq:LkinPhiQ}) the gauge boson mass terms\begin{equation}
g^{2}\sum_{i=1}^{N}\left[v^{2}A_{i\mu}A_{i}^{\mu}+u^{2}(A_{i\mu}-A_{(i+1)\mu})^{2}\right],\label{eq:GaugeMassMatrix}\end{equation}
where the VEVs~$v=\left\langle \Phi_{i}\right\rangle $ and~$u=\left\langle Q_{i}\right\rangle $
have already been inserted. After diagonalization, the mass eigenvalues~$M_{n}$
of the gauge bosons read\begin{equation}
M_{n}^{2}=g^{2}v^{2}+2g^{2}u^{2}\left(1-\cos2\pi\frac{n}{N}\right),\,\,\,\, n=1,\dots,N.\label{eq:GaugeSpec}\end{equation}
This spectrum can be interpreted as follows: For~$n\ll N$ or~$\left|n-N\right|\ll N$
the link fields generate a linear KK spectrum~$\sim n/R=ngu/N$ with
an overall mass scale~$u\sim10^{-2}\,\textrm{eV}$. The fields~$\Phi_{i}$
provide in addition for the gauge bosons a constant kink mass%
\footnote{For deconstructed supersymmetric gauge theories the wave function
profile has been analyzed in Ref.~\cite{Falkowski:2002vc}.%
} of the order~$v\sim10^{2}\,\textrm{GeV}$. In Sec.~\ref{sub:OffsetMass},
we will show that this comparably large kink mass suppresses the resulting
Casimir energy, which would, for bosonic fields, imply a negative
cosmological constant. 

The Lagrangian~$\mathcal{L}_{\textrm{mass}}[\Psi_{i},Q_{i}]$ contains
terms of the type~$Q_{i}^{\dag}\overline{\Psi}_{i\textrm{L}}\Psi_{(i+1)\textrm{R}}+\textrm{h.c.}$
which give after SSB fermion masses of order~$u$. In our 4D model,
a {}``naive'' transverse lattice of the 5D $U(1)$ theory for a
bulk fermion is easily accommodated by taking the mass and mixing
terms to be\begin{equation}
\mathcal{L}_{\textrm{mass}}[\Psi_{i},Q_{i}]=\frac{u}{2}\cdot\sum_{n=1}^{N}\overline{\Psi}_{n\textrm{L}}\left[\frac{Q_{n}^{\dag}}{u}\Psi_{(n+1)\textrm{R}}-\frac{Q_{n-1}}{u}\Psi_{(n-1)\textrm{R}}\right]+\textrm{h.c.},\label{eq:typeIImasses}\end{equation}
which represents the kinetic term of the fermion in the 5th dimension%
\footnote{An explicit fermion mass term can be forbidden by an appropriate discrete
symmetry, see Sec.~\ref{sec:ZeroPointDeconst}.%
}. In the IR, the Lagrangian in Eq.~(\ref{eq:typeIImasses}) generates
identical KK towers for the left- and right-handed states with masses
in the sub-eV range. The exact form of the KK mass spectra of the
fermions will be calculated in Sec.~\ref{sec:ZeroPointDeconst},
where the kinetic term will also be modified to cope with the fermion
doubling problem, which occurs in the naive treatment of the fermions.
As a result of the small inverse lattice spacing~$u$, we will show
in Sec.~\ref{sub:CasimirFermion} that the fermionic Casimir energy
induces a small positive cosmological constant for small~$N$.

It is also possible to interpret the set of scalars~$\Phi_{i}$ as
one 5D massive scalar on a transverse lattice. The corresponding deconstruction
Lagrangian reads\begin{eqnarray}
\mathcal{L} & = & \sum_{i=1}^{N}\left[(D_{\mu}\Phi_{i})^{2}+(m^{2}+M_{b}^{2})\left|\Phi_{i}\right|^{2}+\lambda_{1}\left|\Phi_{i}\right|^{4}\right.\nonumber \\
 & - & \left.M_{b}^{2}\left(\frac{1}{2}\left|\Phi_{i}\right|^{2}+\frac{1}{2}\left|\Phi_{i+1}\right|^{2}-\Phi_{i}\frac{Q_{i}}{u}\Phi_{i+1}^{\dag}+\textrm{h.c.}\right)\right],\label{eq:PhiDecLagr}\end{eqnarray}
where the terms in the last line mimic in the IR the effects of~$-\sum_{i=1}^{N}M_{\textrm{b}}^{2}\left|\Phi_{i}-Q_{i}\Phi_{i+1}/u\right|^{2}\rightarrow\int_{0}^{R}\textrm{d}y\,(D_{5}\Phi)^{2}$.
Here, the scale~$M_{\textrm{b}}$ defines the lattice spacing in
the 5th dimension and the~$Q_{i}$ are taken as link fields. Choosing~$M_{\textrm{b}}^{2}=\mu u\sim(10^{4}-10^{5}\,\textrm{eV})^{2}$,
we recover similar terms as in the potential~$V$ in Eq.~(\ref{eq:seesawpotential}).
The resulting KK mass spectrum corresponds to the one for the gauge
bosons in Eq.~(\ref{eq:GaugeSpec}), but with a compactification
scale of the order~$M_{\textrm{b}}$ and a constant kink mass~$m\sim10^{2}\,\textrm{GeV}$.
The latter one is large enough to significantly suppress the Casimir
effect for the fields~$\Phi_{i}$ in a manner similar to the gauge
bosons. Note that the terms in~$V$ with coefficients~$\lambda_{3},\lambda_{5}$,
and~$\lambda_{6}$ can be neglected with respect to the expression
in Eq.~(\ref{eq:PhiDecLagr}), since they become~$\sim u^{2}v^{2}$.
Furthermore, the term with coefficient~$\lambda_{4}$ must be forbidden
when requiring for the fields~$\Phi_{i}$ only {}``local'' interactions.
In such a case, Eq.~(\ref{eq:PhiDecLagr}) describes very well all
scalar interactions in~$V$ which involve the fields~$\Phi_{i}$.
Note again, that (at the renormalizable level) in a supersymmetric
case~$\lambda_{4}=0$ due to the holomorphy of the superpotential.

\minisec{Embedding into a disk}

\begin{figure}
\begin{center}\includegraphics[%
  bb=220bp 393bp 383bp 547bp,
  clip,
  scale=0.8]{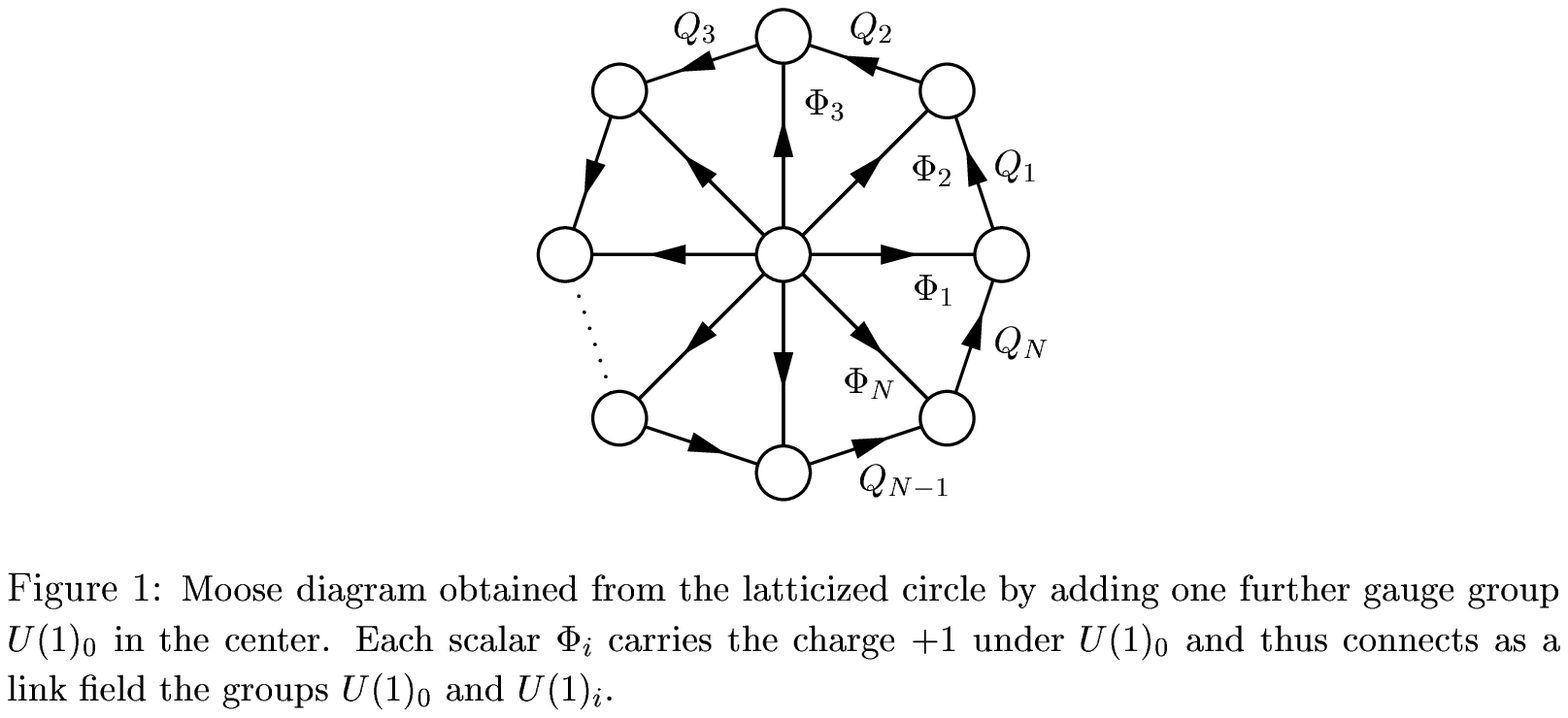}\end{center}

\caption{\label{fig:Plaquette}Moose diagram obtained from the latticized
circle by adding one further gauge group~$U(1)_{0}$ to the center.
Each scalar~$\Phi_{i}$ carries the charge~$+1$ under~$U(1)_{0}$
and thus connects as a link field the groups~$U(1)_{0}$ and~$U(1)_{i}$.}
\end{figure}

So far, we have viewed the fields~$\Phi_{i}$ as scalar site variables.
The potential~$V$, however, possesses a global symmetry which allows
to interpret the fields~$\Phi_{i}$ also as link variables by gauging
the symmetry. In fact, we can embed the latticized 5th dimension into
a disk by adding one further~$U(1)$ gauge group~$U(1)_{0}$ under
which all scalars~$\Phi_{i}$~$(i=1,\ldots,N)$ carry the same charge~$+1$
while the link fields~$Q_{i}$ and the fermions~$\Psi_{i}$ are
all singlets under~$U(1)_{0}$. The moose diagram for these symmetries
is shown in Fig.~\ref{fig:Plaquette}%
 where the extra gauge group~$U(1)_{0}$ has been placed in the center
of a disk whose boundary is the latticized circle discussed above.
Obviously, the addition of the~$U(1)_{0}$ group has promoted the
scalar site variables from the previous model to link fields, since
each~$\Phi_{i}$ is now charged as~$(+1,-1)$ under the product
group~$U(1)_{0}\times U(1)_{i}$. Topological properties of this
lattice geometry have been analyzed in Ref.~\cite{Arkani-Hamed:2001ed}
and its relation to the doublet-triplet splitting problem was addressed
in Ref.~\cite{Witten:2001bf}. It is important to note, that the
extra gauge group~$U(1)_{0}$ does not affect the scalar potential~$V$
and hence the replicated type-II seesaw mechanism which ensures the
smallness of the VEVs~$u=\left\langle Q_{i}\right\rangle \sim10^{-2}\,\textrm{eV}$
remains unaltered. Choosing the gauge coupling of~$U(1)_{0}$ to
be equal to the other~$U(1)_{i}$ gauge couplings~$g$, the Lagrangian
generating the gauge boson masses can be written as \[
\mathcal{L}[\Phi_{i},Q_{i}]=\sum_{i=1}^{N}\left|\left(\partial_{u}-igA_{0\mu}+igA_{i\mu}\right)\Phi_{i}\right|^{2}+|(\partial_{\mu}+igA_{i\mu}-igA_{(i+1)\mu})Q_{i}|^{2}.\]
 In the basis $(A_{0}^{\mu},A_{N}^{\mu},A_{1}^{\mu},A_{2}^{\mu},\ldots,A_{N-1}^{\mu})$,
the~$(N+1)\times(N+1)$ gauge boson mass matrix~$M^{2}$ takes the
form \[
M^{2}=g^{2}v^{2}\left(\begin{matrix}N & -1 & -1 & -1 & \cdots\\
-1 & 1 & 0 & 0 & \cdots\\
-1 & 0 & 1 & 0 & \cdots\\
-1 & 0 & 0 & 1 & \cdots\\
\vdots & \vdots & \vdots & \vdots & \ddots\end{matrix}\right)+g^{2}u^{2}\left(\begin{matrix}0 & 0 & 0 & 0 & \cdots\\
0 & 2 & -1 & 0 & \cdots\\
0 & -1 & 2 & -1 & \cdots\\
0 & 0 & -1 & 2 & \cdots\\
\vdots & \vdots & \vdots & \vdots & \ddots\end{matrix}\right),\]
 where the bottom-right~$N\times N$ submatrix of~$M^{2}$ is just
the gauge boson mass matrix in Eq.~(\ref{eq:GaugeMassMatrix}). Therefore,
the mass eigenstates~$\tilde{A}_{n}^{\mu}$~$(n=1,\dots,N)$ of
this submatrix have masses~$M_{n}$ as in Eq.~(\ref{eq:GaugeSpec}).
Hence,~$M^{2}$ can be brought to diagonal form in two steps. First,
we diagonalize the down-right~$N\times N$ submatrix which gives
in the basis~$(A_{0}^{\mu},\tilde{A}_{N}^{\mu},\tilde{A}_{1}^{\mu},\tilde{A}_{2}^{\mu},\dots,\tilde{A}_{N-1}^{\mu})$
a matrix of the structure \[
M^{2}\rightarrow g^{2}v^{2}\left(\begin{matrix}N & -\sqrt{N} & 0 & 0 & \cdots\\
-\sqrt{N} & 0 & 0 & 0 & \cdots\\
0 & 0 & 0 & 0 & \cdots\\
0 & 0 & 0 & 0 & \cdots\\
\vdots & \vdots & \vdots & \vdots & \ddots\end{matrix}\right)+\left(\begin{matrix}0 & 0 & 0 & 0 & \cdots\\
0 & M_{N}^{2} & 0 & 0 & \cdots\\
0 & 0 & M_{1}^{2} & 0 & \cdots\\
0 & 0 & 0 & M_{2}^{2} & \cdots\\
\vdots & \vdots & \vdots & \vdots & \ddots\end{matrix}\right).\]
 We therefore see, that the rotation in the~$N\times N$ sub-space
has almost diagonalized the total mass matrix. Next, the diagonalization
of~$M^{2}$ is completed by a rotation in the~$(A_{0}^{\mu},\tilde{A}_{N}^{\mu})$
subspace an angle~$\theta=\arctan[\frac{1}{2}(\sqrt{4+(N-1)^{2}/N}-(N-1)/\sqrt{N})]$.
From this we observe, that~$\tilde{A}_{1}^{\mu},\tilde{A}_{2}^{\mu},\dots,\tilde{A}_{N-1}^{\mu}$
are mass eigenstates of~$M^{2}$ since the mixing of these states
with~$A_{0}^{\mu}$ is zero. In other words,~$\tilde{A}_{1}^{\mu},\tilde{A}_{2}^{\mu},\dots,\tilde{A}_{N-1}^{\mu}$
are exactly localized on the boundary of the disk. The remaining two
gauge boson mass eigenstates of~$M^{2}$ are linear combinations
of~$A_{0}^{\mu}$ and~$\tilde{A}_{N}^{\mu}$ which are parameterized
by the mixing angle~$\theta$. In configuration space, the wave function
profiles of these mass eigenstates read~$1/\sqrt{N(N+1)}\cdot(N,-1,-1,\dots,-1)$,
which has mass~$gv\sqrt{N+1}$, and~$1/\sqrt{N+1}\cdot(1,1,\dots,1)$,
which is a zero mode. Obviously, the former massive state is primarily
composed of~$A_{0}^{\mu}$ and becomes arbitrarily well located on
the site in the center for~$N\gg1$. Correspondingly, the flat distribution
of the zero mode over the disk shows that the admixture of~$A_{0}^{\mu}$
to the zero mode vanishes in the limit~$N\rightarrow\infty$. Indeed,
for large~$N$ the mixing angle is approximately~$\theta\sim\sqrt{N}/(N-1)$
and the field~$A_{0}^{\mu}$ {}``decouples'' from the rest of the
gauge bosons such that we effectively recover on the boundary of the
disk a deconstructed 5D~$U(1)$ gauge theory as above. Note that
in~$V$, each tri-linear term~$\propto\mu\Phi_{i}Q_{i}\Phi_{i+1}^{\dag}$
corresponds in Fig.~\ref{fig:Plaquette} to a small triangle and
is therefore interpreted as a gauge-invariant plaquette term with
trivial holonomy in the lowest energy state. The system with real
VEVs as in Eq.~(\ref{eq:universalVEVs}) is gauge-equivalent with
a vacuum structure\[
\left\langle Q_{n}\right\rangle =u\cdot e^{i2\pi/N},\,\,\,\,\left\langle \Phi_{n}\right\rangle =v\cdot e^{i2\pi n/N},\,\,\,\, n=1,\dots,N,\]
which maintains an equivalence between the links on the boundary under
a rotation of the disk by an angle~$2\pi/N$. This rotation yields
a representation of the fundamental group of the boundary, which is~$Z_{N}$~\cite{Witten:2001bf}. 

%\newpage

\section{Casimir Energies and the Cosmological Constant\label{sec:ZeroPointCC}}

The Casimir effect is a notable exception from the normal ordering
procedure in quantum field theories. It occurs when quantum fields
have to obey certain boundary conditions, \textit{e.g.}, the electric
component of the photon field, restricted between two parallel conducting
plates, has to vanish on the plates. This causes a geometry dependent
vacuum energy density inducing a force on the plates. Therefore, the
Casimir effect is a macroscopic quantum phenomenon, which is experimentally
well established~\cite{Sparnaay:1958wg}. For a recent review of
the effect and its applications see, \emph{e.g.,} Ref.~\cite{Bordag:2001qi}.

It has been pointed out, that the Casimir effect is also relevant
in higher-dimensional theories~\cite{Appelquist:1982zs}, where the
bulk fields are subject to boundary conditions associated with a non-trivial
space-time topology. To establish contact with the discussion in Sec.~\ref{sec:Model},
let us from now on restrict our considerations mainly to the Casimir
effect in a 5D~$U(1)$ gauge theory compactified on the circle~$\mathcal{S}^{1}$
with circumference~$R$. After integrating out the extra dimension,
the number of KK modes for each bulk field is typically given by~$N\sim M_{*}R/2\pi$.
In the effective 4D description, each mode contributes, upon quantization,
a divergent amount of zero-point energy to the total vacuum energy
density, which has the form of a cosmological constant. Even for quantum
fields in flat Minkowski space-time, there are always such contributions,
which are usually discarded by normal-ordering. These divergent terms
arise as VEVs of the Hamiltonian density~$\mathcal{H}$, which can
be written as an integral over the energy~$\omega(p)$ of a field
mode with momentum~$p$ and mass~$M$, \emph{i.e.},~$\left\langle \mathcal{H}\right\rangle \propto\int\textrm{d}^{3}p\cdot\omega(p)$
with~$\omega^{2}=\vec{p}^{\,2}+M^{2}$. In the effective 4D description
of the extra-dimensional theory, the total energy density of the~$N$
KK field modes can then be brought into the form \[
\rho\propto\sum_{n=1}^{N}\int\textrm{d}^{3}p\cdot\sqrt{\vec{p}^{\,2}+M^{2}+f(n)},\]
 where the function~$f(n)$ depends on~$N$,~$R$, and the spin
of the fields. Without any knowledge of the 5th dimension, it would
not be clear how to put these UV divergent expressions in a sensible
(finite) form. However, since we can interpret the KK tower in terms
of an underlying higher-dimensional theory with certain boundary conditions,
the 5D Casimir effect provides a well-known procedure to handle these
UV divergences in four dimensions. With this, one obtains an unambiguous
finite expression for the vacuum energy.

Since the UV divergences are all subtracted in the renormalization
program, Casimir energies can be regarded as an IR effect which is
insensitive to the UV details of the theory. We hence expect essentially
similar Casimir energies to emerge in a class of models which have
identical 5D physics as an IR limit but may differ significantly in
the UV. In this section, we shall examine this aspect more properly
by considering the Casimir effect for a 5D theory which is treated
in the UV as a transverse lattice for the extra dimension.%
\footnote{For a related discussion of vacuum energy in a multi-graviton theory
see Ref.~\cite{Kan:2002rp}.%
} In this framework, we will hence be working in a total manifold with
topology~$\mathcal{M}\times\mathcal{S}_{\textrm{lat}}^{1}$, where~$\mathcal{M}$
is the (continuous) Minkowski space and~$\mathcal{S}_{\textrm{lat}}^{1}$
denotes the latticized 5th dimension compactified on the circle. 

In Sec.~\ref{sub:CasimirScalar}, we first discuss the possible field
configurations in~$\mathcal{M}\times\mathcal{S}^{1}$ before we determine
the Casimir energy for a real 5D scalar field~$\phi$ and calculate
the effective cosmological constant in the transverse lattice. The
results for other bosonic fields differ at most by a simple factor,
taking into account the degrees of freedom, \emph{e.g.,} one massive
vector field counts as~$3$ real scalars. Next, in Sec.~\ref{sub:CasimirFermion},
we will perform the analogical calculation for Dirac fermions. Finally,
we briefly summarize in Sec.~\ref{sub:CasimirSummary} our results
for massless fields, and in Sec.~\ref{sub:OffsetMass} we discuss
the effect of a kink mass term.

\subsection{The Casimir effect for scalar fields\label{sub:CasimirScalar}}

At the quantum level, global properties of non-trivial space-time
topology may be probed by the Casimir effect which is sensitive to
the IR structure of a theory. In this context, the existence of inequivalent
field configurations associated with the different boundary conditions
in the space-time manifold is of special interest. The impact of boundary
conditions on the Casimir energies of fields propagating in a non-trivial
space-time becomes already evident for the simple example of a 5D
theory with topology~$\mathcal{M}\times\mathcal{S}^{1}$. For this
case, let us now describe the possible field configurations in the
language of orbifolds%
\footnote{We follow here the treatment of Ref.~\cite{Hebecker:2001jb}.%
}. To this end, we consider a 5D scalar field~$\phi$ defined on the
simply connected manifold~$\mathcal{M}\times\mathbb{R}^{1}$, where
the real line~$\mathbb{R}^{1}$ is described by the coordinate~$y$.
We assume that~$\mathbb{R}^{1}$ and the field~$\phi$ are both
subject to a discrete symmetry group~$K$. Here, we actually take~$K=\mathbb{Z}$,
the additive group of integers. The group~$K$ acts on~$\mathbb{R}^{1}$
as~$K:\, y\rightarrow y+m\cdot R$, where~$m\in K$ and~$R$ is
some finite displacement, while the action of~$K$ on~$\phi$ is~$K:\,\phi\rightarrow D(m)\phi$,
where~$D(m)$ is some (matrix) representation of~$K$. Now, one
can orbifold the theory by requiring the field~$\phi$ to be invariant
under both these actions. Then, the equivalence relation~$y\sim y+R$
imposed by~$K$ on~$\mathbb{R}^{1}$ constrains the true physical
space to be the smooth%
\footnote{This is a result of~$K$ acting \textit{freely}, \textit{i.e.},~$K$
has no fixed points in~$\mathbb{R}^{1}$.%
} manifold~$\mathcal{M}\times\mathbb{R}^{1}/\mathbb{Z}=\mathcal{M}\times\mathcal{S}^{1}$,
where the circle~$\mathcal{S}^{1}$ has the circumference~$R$.
For the simple case of a \emph{real} field~$\phi$, we can take the
matrices~$D(m),\, m\in K$ to form a representation of~$GL(1,\mathbb{R})\sim O(1,\mathbb{R})$
which is just the group~$Z_{2}=\left\{ +1,-1\right\} $. As a result,~$\phi(y=0)$
and~$\phi(y=R)$ are only equal up to a sign, which can also be understood
in the context of fibre bundle theory. In this framework~\cite{Isham:1978xxx},
the field~$\phi$ is interpreted as a cross section%
\footnote{A cross section of a bundle assigns to each point in the base space
a vector in the fibre over that point.%
} of a vector bundle, where the fibre is a real line~$\mathbb{R}$.
Then we have two possibilities to attach the fibre on the base space~$\mathcal{M}\times\mathcal{S}^{1}$.
The first one corresponds to forming a product bundle of the base
space and the real line, implying a cylinder-like structure and thus
periodic boundary conditions for the field, $\phi(y+R)=\phi(y)$.
On the other hand, one can form a non-product bundle by twisting the
fibres which yields a Möbius band and therefore, $\phi$, as a cross
section, obeys anti-periodic boundary conditions\[
\phi(y+mR)=(-1)^{m}\phi(y),\,\,\,\, m\in\mathbb{Z},\]
 because one must cycle twice through the circle~$\mathcal{S}^{1}$
to completely traverse the Möbius band. In the latter case, the field~$\phi$
is called a \emph{twisted} field, whereas cross sections through the
product bundle are \emph{untwisted} fields. Locally, both bundle types
have the same product structure, but globally they differ significantly.
These two bundles are the only ones that can be formed by gluing together
two trivial real vector bundles. They represent the two possibilities
of matching the vector bundles by transition functions, which are
the elements of the structure group, in our case~$Z_{2}$, that acts
on the fibre~$\mathbb{R}^{1}$. Since untwisted and twisted bundles
provide inequivalent cross sections, they yield inequivalent degrees
of freedom of the field~$\phi$, which must be considered in the
Casimir effect.

Turning to the calculation of Casimir energies, we will first treat
the compactified extra dimension to be continuous, before we examine
the lattice description. As above, the position in the extra-dimensional
space is described by the spatial coordinate~$y$ and the corresponding
momentum is called~$q$. For a real bulk-scalar~$\phi$ in the flat
manifold~$\mathcal{M}\times\mathcal{S}^{1}$, it is sensible to use
the plane wave Ansatz\begin{equation}
\phi(t,\vec{x},y)=A\cdot\exp(i\omega t-i\vec{p}\vec{x}-iqy),\label{eq:ScalarFieldAnsatz}\end{equation}
where~$A$ is a normalization factor. As discussed previously, the
untwisted field configuration is fixed by periodic boundary conditions,\[
\phi(y+R)=\phi(y)\,\,\,\,\Longrightarrow\,\,\,\, e^{-iqR}=1,\]
 implying a discrete momentum spectrum:\begin{equation}
q=2\pi\frac{n}{R},\,\,\,\, n\in\mathbb{Z}.\label{eq:QSpectrumUntwist}\end{equation}
For twisted fields we use anti-periodic boundary conditions\[
\phi(y+mR)=(-1)^{m}\phi(y),\,\,\,\, m\in\mathbb{Z}\,\,\,\,\Longrightarrow\,\,\,\, e^{-iqR}=-1,\]
which yield the discrete momentum spectrum\begin{equation}
q=2\pi\frac{(n-\frac{1}{2})}{R},\,\,\,\, n\in\mathbb{Z}.\label{eq:QSpectrumTwist}\end{equation}
Since this is the only difference which is relevant for the following
calculations, we will work with untwisted fields and replace~$n$
by~$n-1/2$ when needed.

\minisec{Discretization}

Taking the latticized nature of the 5th dimension into account, the
discretization of the circle~$\mathcal{S}^{1}$ also forces the coordinate~$y$
to be discrete. Assuming~$N$ lattice sites with a universal lattice
spacing~$a$, the circumference of the 5th dimension is given by~$R=Na$,
and the position~$y$ of each site can be described by a coordinate
index~$j$,\begin{equation}
y=a\cdot j,\,\,\,\, j=1,\dots,N.\label{eq:ScalarDiscreteY}\end{equation}
From the standard definition for a derivative in the continuum,\[
\frac{\partial\phi}{\partial y}(y)=\lim_{a\rightarrow0}\frac{\phi(y+a)-\phi(y)}{a},\]
 follows the discrete forward and backward difference operators~$\partial_{5}\phi$
and~$(\partial_{5}\phi)^{\dag}$:\begin{eqnarray*}
\frac{\partial\phi}{\partial y}\longrightarrow(\partial_{5}\phi) & := & a^{-1}(\phi(j+1)-\phi(j)),\\
\left(\frac{\partial\phi}{\partial y}\right)^{\dag}\longrightarrow(\partial_{5}\phi)^{\dag} & := & a^{-1}(\phi(j)-\phi(j-1)).\end{eqnarray*}
By inserting the Ansatz~(\ref{eq:ScalarFieldAnsatz}) for~$\phi$
we find\[
\partial_{5}\phi=a^{-1}(e^{-iqa}-1)\phi,\,\,\,\,(\partial_{5}\phi)^{\dag}=a^{-1}(1-e^{iqa})\phi,\]
and \[
\partial_{5}^{\dag}\partial_{5}\phi=-2a^{-2}(1-\cos qa)\phi.\]
The Klein-Gordon equation for a real 5D scalar field~$\phi$ with
mass~$M_{\textrm{s}}$ reads\[
\left[\frac{\partial^{2}}{\partial t^{2}}-\nabla^{2}-\partial_{5}^{\dag}\partial_{5}+M_{\textrm{s}}^{2}\right]\phi=0,\]
and determines the energy~$\omega$ of a field mode with the momenta~$\vec{p}$
and~$q$:\[
\omega^{2}=\vec{p}^{\,2}+m^{2},\,\,\,\, m^{2}:=2a^{-2}(1-\cos qa)+M_{\textrm{s}}^{2}.\]

The 5D energy-momentum tensor of the real scalar field~$\phi$ has
the form\begin{equation}
T_{AB}=(\phi_{,A})^{\dag}(\phi_{,B})-\frac{1}{2}g_{AB}g^{CD}(\phi_{,C})^{\dag}(\phi_{,D})+\frac{1}{2}g_{AB}M_{\textrm{s}}^{2}\phi^{\dag}\phi,\label{eq:Scalar5DEngMomTensor}\end{equation}
where~$A,B,C,\dots$ are 5D coordinate indices. Here,~$A=0$ is
the time-like index,~$A=1,2,3$ are spatial indices, corresponding
to the uncompactified $3$-space, and~$A=5$ characterizes the extra
spatial dimension. The 5D energy density~$\rho_{5}$ is the $00$-component
of~$T_{AB}$:\begin{equation}
\rho_{5}=T_{00}=\frac{1}{2}(\partial_{0}\phi)^{\dag}(\partial_{0}\phi)+\frac{1}{2}(\nabla\phi)^{\dag}(\nabla\phi)+\frac{1}{2}(\partial_{5}\phi)^{\dag}(\partial_{5}\phi)+\frac{1}{2}M_{\textrm{s}}^{2}\phi^{\dag}\phi.\label{eq:EngDensClass}\end{equation}
 By averaging over all directions of the isotropic $3$-space, we
obtain the pressure~$p_{5}$ of the scalar field~$\phi$: \begin{equation}
p_{5}=\frac{1}{3}\sum_{i=1}^{3}T_{ii}=\frac{1}{3}\left[(\nabla\phi)^{\dag}(\nabla\phi)+\frac{1}{2}(\partial_{0}\phi)^{\dag}(\partial_{0}\phi)-\frac{1}{2}(\nabla\phi)^{\dag}(\nabla\phi)-\frac{1}{2}(\partial_{5}\phi)^{\dag}(\partial_{5}\phi)-\frac{1}{2}M_{\textrm{s}}^{2}\phi^{\dag}\phi\right].\label{eq:PressClass}\end{equation}
The (canonical) quantization of the field~$\phi$ leads to the field
operator\begin{equation}
\hat{\phi}(t,\vec{x},y)=\sqrt{V_{3}}\int\textrm{d}^{3}p\sum_{n=1}^{N}\left(\phi(\vec{p},n)a_{\vec{p},n}+\phi^{\dag}(\vec{p},n)a_{\vec{p},n}^{\dag}\right),\label{eq:ScalarFeldop}\end{equation}
where~$a_{p,n}$ and~$a_{p,n}^{\dag}$ obey the bosonic commutator
relations\[
[a_{\vec{p},n},a_{\vec{p}^{\,\prime},n^{\prime}}^{\dag}]=V_{3}^{-1}\delta(\vec{p}-\vec{p}^{\,\prime})\delta_{nn^{\prime}}\,,\,\,\,\,[a,a]=[a^{\dag},a^{\dag}]=0.\]
The energy-momentum operator~$\hat{T}_{AB}$ is obtained by replacing~$\phi$
in Eq.~(\ref{eq:Scalar5DEngMomTensor}) by the field operator~$\hat{\phi}$.
After some calculation (see Appendix~\ref{secA:ScalarEngPress}),
we arrive at the energy density $\rho_{5}$ and the pressure $p_{5}$
of the 5D quantized field $\phi$,\begin{eqnarray}
\rho_{5}=\left\langle 0\right|\hat{T}_{00}\left|0\right\rangle  & = & \frac{1}{2(2\pi)^{3}R}\int\textrm{d}^{3}p\sum_{n=1}^{N}\,\omega,\label{eq:ScalarRhoUnregul}\\
p_{5}=\frac{1}{3}\sum_{i=1}^{3}\left\langle 0\right|\hat{T}_{ii}\left|0\right\rangle  & = & \frac{1}{2(2\pi)^{3}R}\int\textrm{d}^{3}p\sum_{n=1}^{N}\,\frac{\vec{p}^{\,2}}{3\omega}.\label{eq:ScalarPUnregul}\end{eqnarray}

\minisec{Regularization}

The momentum integral~$\int\textrm{d}^{3}p=4\pi\int_{0}^{\infty}\textrm{d}p\cdot p^{2}$
in Eqs.~(\ref{eq:ScalarRhoUnregul}) and~(\ref{eq:ScalarPUnregul})
and thus~$\rho_{5}$ and~$p_{5}$ are divergent. We introduce therefore
a regularization to obtain meaningful, finite expressions. Consider
first Eq.~(\ref{eq:ScalarRhoUnregul}). Introducing an exponential
suppression factor~$e^{-(dp)^{2}}$ as regulator function for~$d>0$,
we have\begin{eqnarray}
\int\textrm{d}p\cdot p^{2}\omega & \longrightarrow & \int_{0}^{\infty}\textrm{d}p\cdot p^{2}\omega e^{-(dp)^{2}}\nonumber \\
 & = & \frac{1}{4}m^{2}d^{-2}e^{\frac{1}{2}(dm)^{2}}K_{1}\left(\kl{\frac{1}{2}}d^{2}m^{2}\right)\nonumber \\
 & = & \frac{1}{2}d^{-4}+\frac{1}{4}m^{2}d^{-2}+\frac{1}{8}m^{4}\left(\frac{1}{4}+\frac{1}{2}\gamma-\ln2+\ln(dm)\right)+\mathcal{O}(d^{6}m^{6}),\label{eq:ScalarRhoRegul}\end{eqnarray}
where~$K_{n}(x)$ is the modified Bessel function of the second kind
of the order~$n$ and~$\gamma=0,577\dots$ is the Euler-Mascheroni
constant. The limit~$d\rightarrow0$ removes the regulator and recovers
the divergence. Before taking that limit, a renormalization has to
be carried out to remove the potentially divergent terms. Alternatively
to the exponential regulator function in Eq.~(\ref{eq:ScalarRhoRegul}),
one can also apply dimensional regularization by moving to~$n$ space-time
dimensions. To be specific, \begin{eqnarray*}
\int\textrm{d}^{3}p\cdot\omega & \longrightarrow & \mu^{(4-n)}S(n-1)\cdot\int_{0}^{\infty}\textrm{d}p\cdot p^{(n-2)}\cdot\omega\\
 & = & -\frac{1}{2}m^{4}\left(\frac{\mu}{m}\right)^{(4-n)}\pi^{(n/2-1)}\Gamma\left(-\frac{n}{2}\right)\\
 & \stackrel{n\rightarrow4}{=} & (4\pi)\frac{1}{8}m^{4}\left[(n-4)^{-1}+\frac{1}{2}\gamma+\ln\left(\frac{m}{\mu}\right)+\mathcal{O}(n-4)\right]_{n\rightarrow4},\end{eqnarray*}
where the arbitrary energy scale~$\mu$ has been introduced to keep
the dimension of the whole term constant, and~$S(n)$ is the surface
area of an $n$-ball. 

For a curved background space-time~\cite{Birrell:ix}, one would
decompose~$\rho$ into a divergent and a finite term, so that the
former one has the form of a cosmological term in Einsteins´s equations.
Then the divergences would be absorbed into renormalized coupling
constants (like~$\Lambda$), and the finite remainder is called the
\emph{renormalized} energy density or, in our case, the Casimir energy
density. Here, such a general treatment is not necessary because the
divergence also arises in flat space-time, like our~$\mathcal{M}\times\mathcal{S}^{1}$-manifold,
but there are neither cosmological terms nor Einstein´s equations.
In order to get rid of the divergence, one simply subtracts the corresponding
part of the energy density of the same field in a Minkowski-like space-time
with the same dimensions, \emph{i.e.},~$\mathcal{M}\times\mathbb{R}^{1}$
in our case. This kind of renormalization works because the 5D Minkowski
space suffers from the same divergence as the~$\mathcal{M}\times\mathcal{S}^{1}$
space-time but exhibits no Casimir effect. 

Before discussing the details of the renormalization process, we determine
the regularized form of the pressure~$p$ of the 5D quantized field~$\phi$.
The regularization with the exponential suppression factor with~$d>0$
goes along the same lines as above:\begin{eqnarray}
\int\textrm{d}p\cdot p^{2}\frac{p^{2}}{\omega} & \longrightarrow & \int_{0}^{\infty}\textrm{d}p\cdot p^{2}\frac{p^{2}}{\omega}e^{-(dp)^{2}}\nonumber \\
 & = & \frac{1}{4}m^{2}d^{-2}e^{\frac{1}{2}(dm)^{2}}\left(d^{2}m^{2}K_{0}\left(\kl{\frac{1}{2}}d^{2}m^{2}\right)+(1-d^{2}m^{2})K_{1}\left(\kl{\frac{1}{2}}d^{2}m^{2}\right)\right)\nonumber \\
 & = & \frac{1}{2}d^{-4}-\frac{1}{4}m^{2}d^{-2}-\frac{3}{8}m^{4}\left(\frac{7}{12}+\frac{\gamma}{2}-\ln2+\ln(dm)\right)+\mathcal{O}(d^{6}m^{6}).\label{eq:ScalarPRegul}\end{eqnarray}
Notice that when keeping in in Eqs.~(\ref{eq:ScalarRhoRegul}) and~(\ref{eq:ScalarPRegul})
only terms proportional to~$m^{4}\ln m$, we obtain an equation of
state~$p=-\rho$ of a cosmological constant. As we will see next,
the calculation of the Casimir effect on the lattice indeed involves
for~$d\rightarrow0$ an exact elimination of all terms which are
different from~$m^{4}\ln m$ thus leaving a finite contribution to
the cosmological constant.

\minisec{Renormalization}

The discrete mode sums for the energy density and pressure in Eqs.~(\ref{eq:ScalarRhoUnregul})
and~(\ref{eq:ScalarPUnregul}) are a result of imposing the periodic
or anti-periodic boundary conditions on the lattice scalar. For a
quantized scalar field~$\hat{\phi}$ which lives on the lattice in
the same volume of space (with length~$R$ in the fifth dimension)
and which does not obey such boundary conditions, the energy density
and pressure can be written as\begin{eqnarray}
\rho_{5} & = & \frac{1}{2(2\pi)^{3}R}\int\textrm{d}^{3}p\cdot R\int_{0}^{2\pi/a}\frac{\textrm{d}q}{2\pi}\omega=\frac{1}{2(2\pi)^{3}R}\int\textrm{d}^{3}p\cdot N\int_{0}^{1}\textrm{d}s\cdot\omega,\label{eq:ScalarRho5Free}\\
p_{5} & = & \frac{1}{2(2\pi)^{3}R}\int\textrm{d}^{3}p\cdot R\int_{0}^{2\pi/a}\frac{\textrm{d}q}{2\pi}\frac{\vec{p}^{\,2}}{3\omega}=\frac{1}{2(2\pi)^{3}R}\int\textrm{d}^{3}p\cdot N\int_{0}^{1}\textrm{d}s\frac{\vec{p}^{\,2}}{3\omega},\label{eq:ScalarP5Free}\end{eqnarray}
where~$\omega^{2}=\vec{p}^{\,2}+2a^{-2}(1-\cos qa)+M_{\textrm{s}}^{2}$.
In Appendix~\ref{secA:Renormalization}, it is explicitly shown how
these mode integrals correspond to the uncompactified latticized space.
To calculate the Casimir energy density in~$\mathcal{M}\times\mathcal{S}_{\textrm{lat}}^{1}$,
we subtract from the energy density (\ref{eq:ScalarRhoUnregul}) of
the field subject to the boundary conditions the corresponding energy
density (\ref{eq:ScalarRho5Free}) of the field without boundary conditions,
which gives the renormalized energy density\begin{eqnarray}
\rho_{5} & = & \frac{1}{2(2\pi)^{3}R}\int\textrm{d}^{3}p\left[\sum_{n=1}^{N}\omega-N\int_{0}^{1}\textrm{d}s\omega\right]\nonumber \\
 & = & \frac{1}{2(2\pi)^{3}R}\cdot\frac{4\pi}{8}\left[\sum_{n=1}^{N}m^{4}\ln m-N\cdot\int_{0}^{1}\textrm{d}s\cdot m^{4}\ln m\right]\nonumber \\
 & = & +R^{-5}S_{1}(N),\,\,\,\,\textrm{(untwisted)}\label{eq:ScalarRho5Ren}\end{eqnarray}
where~$m^{2}=2a^{-2}(1-\cos qa)+M_{\textrm{s}}^{2}$ and where we
have introduced the function\begin{eqnarray}
S_{1}(N) & := & \frac{1}{4(2\pi)^{2}}N^{4}\cdot\left[\sum_{n=1}^{N}\left(1-\cos2\pi\frac{n}{N}+\frac{1}{2}a^{2}M_{\textrm{s}}^{2}\right)^{2}\ln\left(1-\cos2\pi\frac{n}{N}+\frac{1}{2}a^{2}M_{\textrm{s}}^{2}\right)\right.\nonumber \\
 & - & \left.N\cdot\int_{0}^{1}\textrm{d}s\cdot\left(1-\cos2\pi s+\frac{1}{2}a^{2}M_{\textrm{s}}^{2}\right)^{2}\ln\left(1-\cos2\pi s+\frac{1}{2}a^{2}M_{\textrm{s}}^{2}\right)\right].\label{eq:CasimirFuncS1}\end{eqnarray}
In Eq.~(\ref{eq:ScalarRho5Ren}) we have first regulated the expressions
with the exponential regulator method in Eq.~(\ref{eq:ScalarRhoRegul})
before applying the relations as given in Appendix \ref{secA:Renormalization}
to exactly subtract for~$d\rightarrow0$ all terms of the type\[
\frac{1}{2}d^{-4}+\frac{1}{4}m^{2}d^{-2}+\frac{1}{8}m^{4}\left(\frac{1}{4}+\frac{1}{2}\gamma-\ln2+\ln(d)\right)+\mathcal{O}(d^{6}m^{6}).\]
For the renormalized pressure of the quantized field, we apply the
corresponding subtraction as in Eq.~(\ref{eq:ScalarRho5Ren}), \emph{i.e.},\begin{eqnarray}
p_{5} & = & \frac{1}{2(2\pi)^{3}R}\int\textrm{d}^{3}p\left[\sum_{n=1}^{N}\frac{\vec{p}^{\,2}}{3\omega}-N\int_{0}^{1}\textrm{d}s\frac{\vec{p}^{\,2}}{3\omega}\right]\nonumber \\
 & = & \frac{-1}{2(2\pi)^{3}R}\cdot\frac{4\pi}{8}\left[\sum_{n=1}^{N}m^{4}\ln m-N\cdot\int_{0}^{1}\textrm{d}s\cdot m^{4}\ln m\right]\nonumber \\
 & = & -R^{-5}S_{1}(N),\,\,\,\,\textrm{(untwisted)}\label{eq:ScalarP5Ren}\end{eqnarray}
where we again regulated the divergent expressions following Eq.~(\ref{eq:ScalarPRegul})
and then exactly subtracted for~$d\rightarrow0$ all terms of the
form\[
\frac{1}{2}d^{-4}-\frac{1}{4}m^{2}d^{-2}-\frac{3}{8}m^{4}\left(\frac{7}{12}+\frac{1}{2}\gamma-\ln2+\ln(d)\right)+\mathcal{O}(d^{6}m^{6}).\]
Comparision of Eqs.~(\ref{eq:ScalarRho5Ren}) and~(\ref{eq:ScalarP5Ren})
shows that the renormalized finite values of~$\rho_{5}$ and~$p_{5}$
obey an equation of state~$p=-\rho$ which is that of a cosmological
constant and hence, the renormalization precedure carried out here
actually amounts to a renormalization of the effective cosmological
constant in the 4D subspace. As a matter of fact, it is sufficient
to restrict in the following our considerations to the vacuum energy
density alone with corresponding statements for the pressure implied.

In the limit~$N\rightarrow\infty$ and~$M_{\textrm{s}}=0$, the
function~$S_{1}(N)$ converges to the value of a continuous 5th dimension:\[
\lim_{N\rightarrow\infty}S_{1}(N)=-\frac{1}{4(2\pi)^{2}}3\zeta(5)\cdot4.\]
By integrating out the 5th dimension, we obtain the 4D energy density
\begin{equation}
\rho_{4}=\int_{0}^{R}\textrm{d}r\cdot\rho_{5}=R\cdot\rho_{5}=-\frac{3\zeta(5)}{(2\pi)^{2}R^{4}}=\frac{1}{R^{4}}\cdot(-0,0787970\dots).\,\,\,\,(\textrm{untwisted})\label{eq:ScalarR4UntwistErg}\end{equation}

In the case of a twisted scalar field everything is like above, but
the energy density reads~$\rho_{5}=+R^{-5}S_{2}(N)$, where~$S_{2}(N)$
is the function~$S_{1}(N)$ with~$n$ replaced by~$n-\frac{1}{2}$.
For massless fields~($M_{\textrm{s}}=0$) we obtain in the continuum
limit\[
\lim_{N\rightarrow\infty}S_{2}(N)=+\frac{1}{4(2\pi)^{2}}3\zeta(5)\cdot(4-\kl{\frac{1}{4}}),\]
and after integrating out the 5th dimension the 4D energy density
reads\begin{equation}
\rho_{4}=R\cdot\rho_{5}=+\frac{15}{16}\cdot\frac{3\zeta(5)}{(2\pi)^{2}R^{4}}=\frac{1}{R^{4}}\cdot(+0,0738722\dots).\,\,\,\,(\textrm{twisted})\label{eq:ScalarR4TwistErg}\end{equation}
Obviously, untwisted and twisted fields provide energy densities~$\rho_{4}$
of different sign. Note that the values for~$\rho_{4}$ in Eqs.~(\ref{eq:ScalarR4UntwistErg})
and~(\ref{eq:ScalarR4TwistErg}) agree with the results in Refs.~\cite{Candelas:ae,Kantowski:ct}.

\subsection{The Casimir effect for fermions\label{sub:CasimirFermion}}

In analogy with the treatment of scalar fields in Sec.~\ref{sub:CasimirScalar},
we will now calculate the Casimir energy density of Dirac fermions.
Therefore, a plane wave Ansatz for Dirac spinor fields~$\Psi$ in
the~$\mathcal{M}\times\mathcal{S}^{1}$ manifold is a convenient
choice, too: \begin{equation}
\Psi=\psi\exp(i\omega t-i\vec{p}\vec{x}-iqy).\label{eq:FermionFeldAnsatz}\end{equation}
The boundary conditions, associated with the compactified~$\mathcal{S}^{1}$-dimension,
provide the discrete momentum spectra. For twisted and untwisted fields
we have~$q=2\pi(n-\frac{1}{2})/(aN)$ and~$q=2\pi n/(aN)$, respectively.
Like in Eq.~(\ref{eq:ScalarDiscreteY}) the coordinate~$y$ corresponding
to the 5th dimension is discrete,~$y=a\cdot j$, where~$j=1,\dots,N$,
and implies an upper bound for the momentum~$q$.

\begin{figure}[t]
\begin{center}\includegraphics[%
  clip,
  width=0.95\textwidth,
  keepaspectratio]{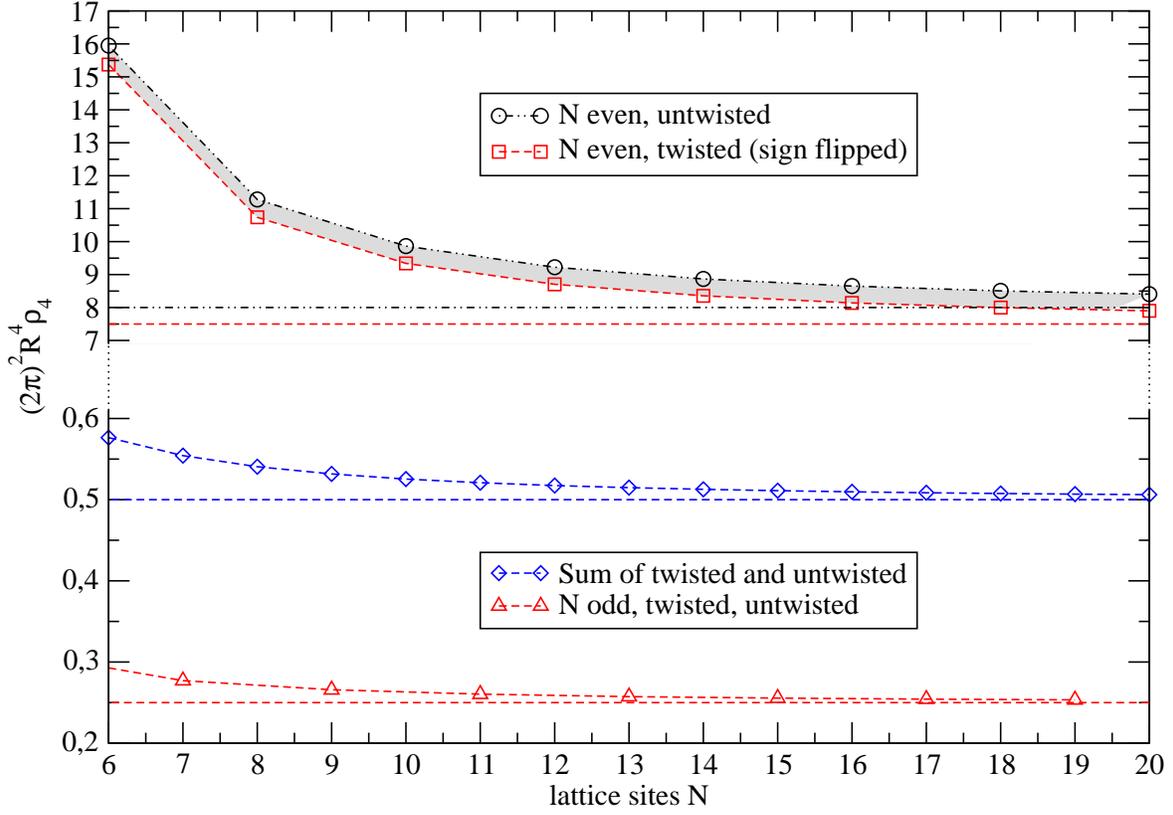}\end{center}

\caption{\label{fig:OddEvenArtifact}Illustration of the odd-even artefact
for fermion fields. The cases of odd and even~$N$ are plotted separately,
and the circumference~$R$ of the 5th dimension is kept constant.
For a better representation, we flipped the sign in the results for
twisted fields and even~$N$. The horizontal lines correspond to
the continuum values given in Table~\ref{tab:Rho4NInfty}.}
\end{figure}

Unlike the Klein-Gordon equation for scalars fields, the Dirac equation
is linear in the derivatives, and therefore we need a symmetric derivative
operator for the discrete~$y$-coordinate:\[
\partial_{5}\Psi(j):=\frac{1}{2a}\left(\Psi(j+1)-\Psi(j-1)\right).\]
With the Ansatz~(\ref{eq:FermionFeldAnsatz}) we obtain\begin{eqnarray*}
\partial_{5}\Psi(j) & = & \frac{1}{2a}Nu\cdot\exp(-i\omega t+i\vec{p}\vec{x})\left[\exp(iqa(j+1))-\exp(iqa(j-1))\right]\\
 & = & \Psi(j)\cdot\left(+\frac{i}{a}\sin(qa)\right),\end{eqnarray*}
and together with the 5D Dirac equation%
\footnote{A 5th Dirac matrix $\gamma^{5}:=i\gamma_{\textrm{4D}}^{5}$ has to
be introduced, where $\gamma_{\textrm{4D}}^{5}$ is the usual $\gamma^{5}$
matrix of the 4D Dirac theory \cite{Pilaftsis:1999jk}.%
} for a Dirac field with mass~$M_{\textrm{f}}$, \[
(i\gamma^{A}\partial_{A}-M_{\textrm{f}})\Psi=0,\,\,\,\, A=0,\dots,3,5,\]
 the energy-momentum relation is determined to be \begin{equation}
\omega^{2}=\vec{p}^{\,2}+m^{2},\,\,\,\, m^{2}:=a^{-2}\sin^{2}qa+M_{\textrm{f}}^{2}.\label{eq:FermionEngMomRel}\end{equation}
The energy-momentum tensor~$T_{AB}$ for the Dirac field~$\Psi$
has the form\[
T_{AB}=\frac{1}{4}i[\overline{\Psi}\gamma_{A}\partial_{B}\Psi+\overline{\Psi}\gamma_{B}\partial_{A}\Psi-\overline{(\partial_{A}\Psi)}\gamma_{B}\Psi-\overline{(\partial_{B}\Psi)}\gamma_{A}\Psi],\]
and the usual canonical quantization procedure parallels that for
scalar fields up to replacing the bosonic commutator relations by
the fermionic anti-commutator relations, which give an overall minus
sign in the result. The Dirac fermion also has four times the degrees
of freedoms of a real scalar, describing particles and anti-particles
with two spin states each. In total, the energy density~$\rho_{5}$
and pressure~$p_{5}$ of a quantized Dirac field differ from the
scalar results of Eqs.~(\ref{eq:ScalarRhoUnregul}) and~(\ref{eq:ScalarPUnregul})
only by a factor of~$(-4)$ and in the modified energy-momentum relation
of Eq.~(\ref{eq:FermionEngMomRel}), \emph{i.e.},\begin{eqnarray}
\rho_{5} & = & -\frac{2}{(2\pi)^{3}R}\sum_{n=1}^{N}\int\textrm{d}^{3}p\cdot\omega,\label{eq:FermionRhoUnregul}\\
p_{5} & = & -\frac{2}{(2\pi)^{3}R}\sum_{n=1}^{N}\int\textrm{d}^{3}p\cdot\frac{\vec{p}^{\,2}}{\omega},\label{eq:FermionPUnregul}\end{eqnarray}
where~$\omega^{2}=\vec{p}^{\,2}+a^{-2}\sin^{2}qa+M_{\textrm{f}}^{2}$.
From here on, the regularization and renormalization procedures are
identical to the scalar case in Sec.~\ref{sub:CasimirScalar}. This
also implies that the equation of state of the fermionic vacuum energy
is that of a cosmological constant,~$p=-\rho$. Thus, it is sufficient
to give the renormalized energy density in five dimensions \begin{eqnarray*}
\rho_{5} & = & -\frac{2}{(2\pi)^{3}R}4\pi\frac{1}{8}\left[\sum_{n=1}^{N}m^{4}\ln m-N\cdot\int_{0}^{1}\textrm{d}s\cdot m^{4}\ln m\right]\\
 & = & +R^{-5}F_{1}(N),\end{eqnarray*}
where the function~$F_{1}(N)$ in the last equation is defined as\begin{eqnarray}
F_{1}(N) & := & -\frac{1}{4(2\pi)^{2}}N^{4}\cdot\left[\sum_{n=1}^{N}\left(\sin^{2}2\pi\frac{n}{N}+a^{2}M_{\textrm{f}}^{2}\right)^{2}\ln\left(\sin^{2}2\pi\frac{n}{N}+a^{2}M_{\textrm{f}}^{2}\right)\right.\nonumber \\
 & - & \left.N\cdot\int_{0}^{1}\textrm{d}s\cdot\left(\sin^{2}2\pi\frac{n}{N}+a^{2}M_{\textrm{f}}^{2}\right)^{2}\ln\left(\sin^{2}2\pi\frac{n}{N}+a^{2}M_{\textrm{f}}^{2}\right)\right],\label{eq:CasimirFuncF1}\end{eqnarray}
with~$R=Na$. For the twisted Dirac field we have \[
\rho_{5}=R^{-5}F_{2}(N),\]
where~$F_{2}(N)$ is the function~$F_{1}(N)$ with~$n$ replaced
by~$n-\frac{1}{2}$. Unlike the functions~$S_{1,2}(N)$ for the
scalar fields, the functions~$F_{1,2}(N)$ for the fermionic fields
have two limit points each, which depend on whether the number of
lattice sites~$N$ is even or odd. For massless fermions~($M_{\textrm{f}}=0$)
and even~$N$ we obtain\begin{eqnarray*}
\lim_{N\rightarrow\infty}F_{1}(N) & = & -\frac{1}{4(2\pi)^{2}}\cdot3\zeta(5)\cdot(-32),\,\,\,\,(\textrm{untwisted})\\
\lim_{N\rightarrow\infty}F_{2}(N) & = & -\frac{1}{4(2\pi)^{2}}\cdot3\zeta(5)\cdot(+30).\,\,\,\,(\textrm{twisted})\end{eqnarray*}
After integrating out the 5th dimension, the 4D Casimir energy densities
read\begin{eqnarray}
\rho_{4} & = & \frac{32\cdot3\zeta(5)}{4(2\pi)^{2}R^{4}}=\frac{1}{R^{4}}\cdot(+0,630376\dots),\,\,\,\,(\textrm{untwisted})\label{eq:FermR4UntwistErg}\\
\rho_{4} & = & \frac{-30\cdot3\zeta(5)}{4(2\pi)^{2}R^{4}}=\frac{1}{R^{4}}\cdot(-0,590978\dots).\,\,\,\,(\textrm{twisted})\label{eq:FermR4TwistErg}\end{eqnarray}
In the case of odd~$N$, both functions have the same limit\[
\lim_{N\rightarrow\infty}F_{1}(N)=\lim_{N\rightarrow\infty}F_{2}(N)=\frac{1}{4(2\pi)^{2}}\cdot3\zeta(5),\]
\[
\rho_{4}=\frac{3\zeta(5)}{4(2\pi)^{2}R^{4}}=\frac{1}{R^{4}}\cdot(+0,019699\dots).\]
This behavior seems to be an effect of the lattice (\emph{odd-even
artefact}~\cite{Hill:2002me}), since the analytic calculation of
Sec.~\ref{sec:CasimirAnalytic} yields the same values as for even~$N$.
We also notice, that the~$N\rightarrow\infty$ limit of the sum of
twisted and untwisted results does not depend on whether~$N$ is
even or odd. Therefore, it is reasonable to consider only this sum
as a physical quantity. For finite~$N$, this odd-even artefact is
illustrated in Fig.~\ref{fig:OddEvenArtifact}. Note again, that
the results for~$\rho_{4}$ in Eqs.~(\ref{eq:FermR4UntwistErg})
and~(\ref{eq:FermR4TwistErg}) are identical with the values in Ref.~\cite{Candelas:ae}.

%\newpage

\subsection{Summary for massless fields\label{sub:CasimirSummary}}

The calculations of Sec.~\ref{sub:CasimirScalar} show that the Casimir
effect for a real scalar field in the transverse lattice space-time~$\mathcal{M}\times\mathcal{S}_{\textrm{lat}}^{1}$
induces a negative vacuum energy density~$\rho$ and therefore a
negative effective cosmological constant in the 4D subspace. On the
other hand, the fermionic Dirac field of Sec.~\ref{sub:CasimirFermion}
yields a positive contribution to the cosmological constant.

\begin{figure}[t]
\begin{center}\includegraphics[%
  clip,
  width=1.0\textwidth,
  keepaspectratio]{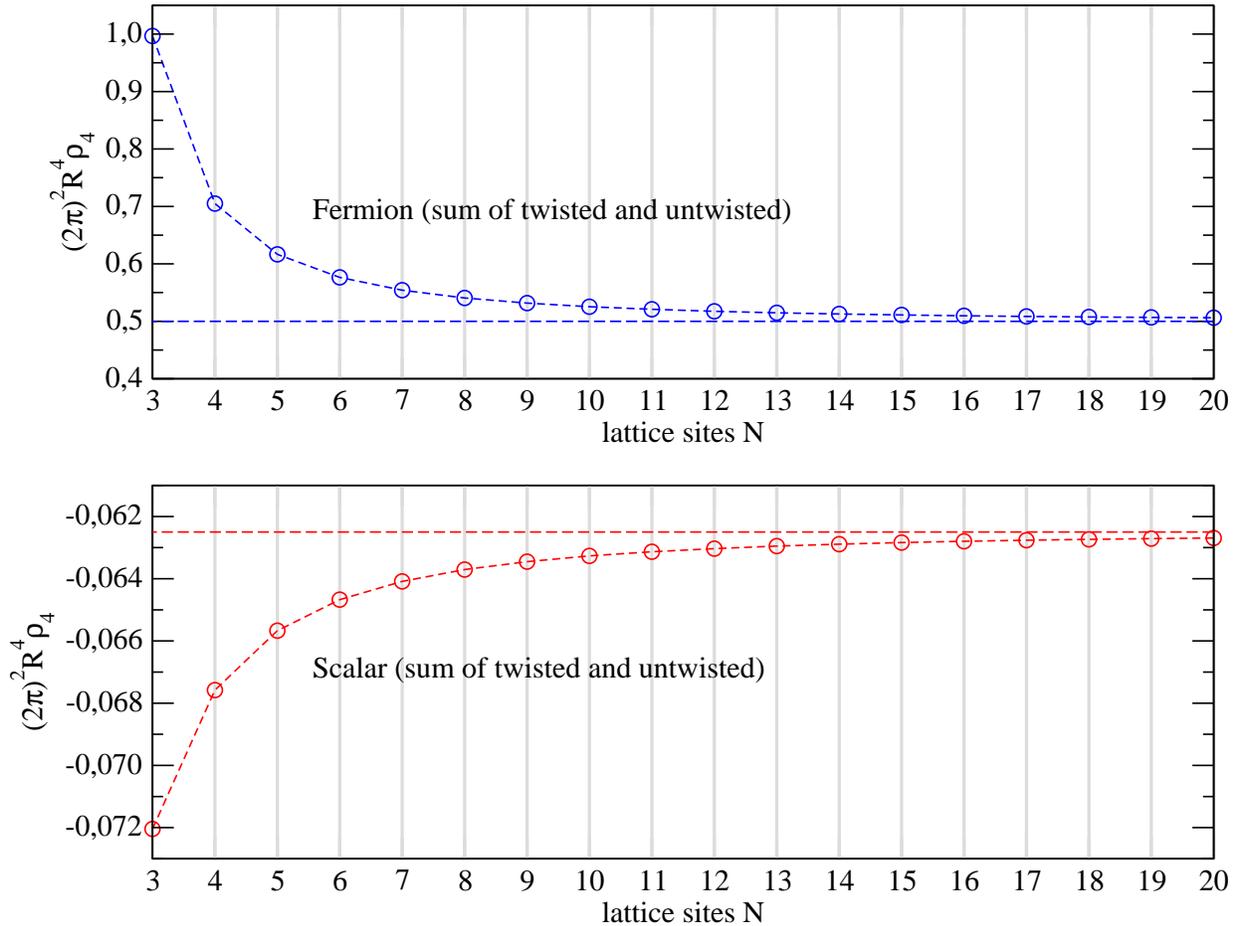}\end{center}

\caption{\label{fig:Rho4ofNScalarFermion}The Casimir energy densities for
fermions (upper graph) and real scalars (lower graph). In this plot,~$R$
is fixed so that the lattice spacing decreases for increasing~$N$.
The dashed horizontal lines denote the continuum limit~$N\rightarrow\infty$,
which has for scalars the value~$-\frac{1}{16}$ and for fermions~$+\frac{1}{2}$.}
\end{figure}

We have already concluded that only the sum of twisted and untwisted
fields can be regarded as a physical quantity, and we note that its
sign is independent of~$N$. Moreover, for a constant circumference~$R$
and small~$N$, the Casimir energy density~$\rho_{4}(N)$ in the
transverse lattice setup has already the same order of magnitude as
the energy density~$\rho_{4}(N\rightarrow\infty)$ in the continuum
limit. Specifically, for~$N\gtrsim10$ the continuum result is approximated
at the few percent level. Even for a number of lattice sites which
is as small as~$N=3$, the results differ at most by a factor of~$2$,
which is clearly shown in Fig.~\ref{fig:Rho4ofNScalarFermion}. 

In the limit~$N\rightarrow\infty$~(Table~\ref{tab:Rho4NInfty}),
the results for real scalars are the same as in the non-lattice calculation~\cite{Candelas:ae,Kantowski:ct},
but for the fermions there is an extra factor of~$2$ in the energy
density of our lattice calculation because of the fermion doubling
phenomenon in lattice theory. In a calculation for continuous dimensions,
one usually expects, from counting degrees of freedom, that the energy
density for Dirac fermions is~$(-4)$~times the value of real scalars. 

Up to now, we have investigated the Casimir effect for Dirac fermions
and real scalars having twisted and untwisted field configurations.
When passing to a complex scalar field which transforms under a~$U(1)$
gauge group there exist only trivial (untwisted) structures and therefore
the charged scalar obeys only periodic boundary conditions. For fermions,
on the other hand, the appearance of twisted field modes is related
to the double covering map~$SL(2,\mathbb{C})\rightarrow SO(3,1)$
which gives rise to inequivalent spin connections~\cite{Isham:1978xxx}.
Consequently, even in presence of a simply connected gauge group like~$U(1)$
we still have also the anti-periodic boundary condition for the fermions.

\begin{table}[t]
\begin{center}\begin{tabular}{|c|c|c||c|}
\hline 
$\rho_{4}R^{4}$&
untwisted&
twisted&
sum\tabularnewline
\hline
\hline 
real scalar&
$-1\cdot(2\pi)^{-2}3\zeta(5)$&
$+\frac{15}{16}\cdot(2\pi)^{-2}3\zeta(5)$&
$-\frac{1}{16}\cdot(2\pi)^{-2}3\zeta(5)$\tabularnewline
\hline 
fermion, $N$ even&
$+8\cdot(2\pi)^{-2}3\zeta(5)$&
$-\frac{15}{2}\cdot(2\pi)^{-2}3\zeta(5)$&
$+\frac{1}{2}\cdot(2\pi)^{-2}3\zeta(5)$\tabularnewline
\hline
fermion, $N$ odd&
$+\frac{1}{4}\cdot(2\pi)^{-2}3\zeta(5)$&
$+\frac{1}{4}\cdot(2\pi)^{-2}3\zeta(5)$&
$+\frac{1}{2}\cdot(2\pi)^{-2}3\zeta(5)$\tabularnewline
\hline
\end{tabular}\end{center}

\caption{\label{tab:Rho4NInfty}The Casimir energy density~$\rho_{4}$ multiplied
by~$R^{4}$ for real massless scalars and Dirac fermions ($M_{\textrm{s,f}}=0$)
in the limit of an infinite number of lattice sites ($N\rightarrow\infty$). }
\end{table}

%\newpage

\subsection{Exponential suppression by a kink mass\label{sub:OffsetMass}}

\begin{figure}[p]
\begin{center}\includegraphics[%
  clip,
  width=0.95\textwidth,
  keepaspectratio]{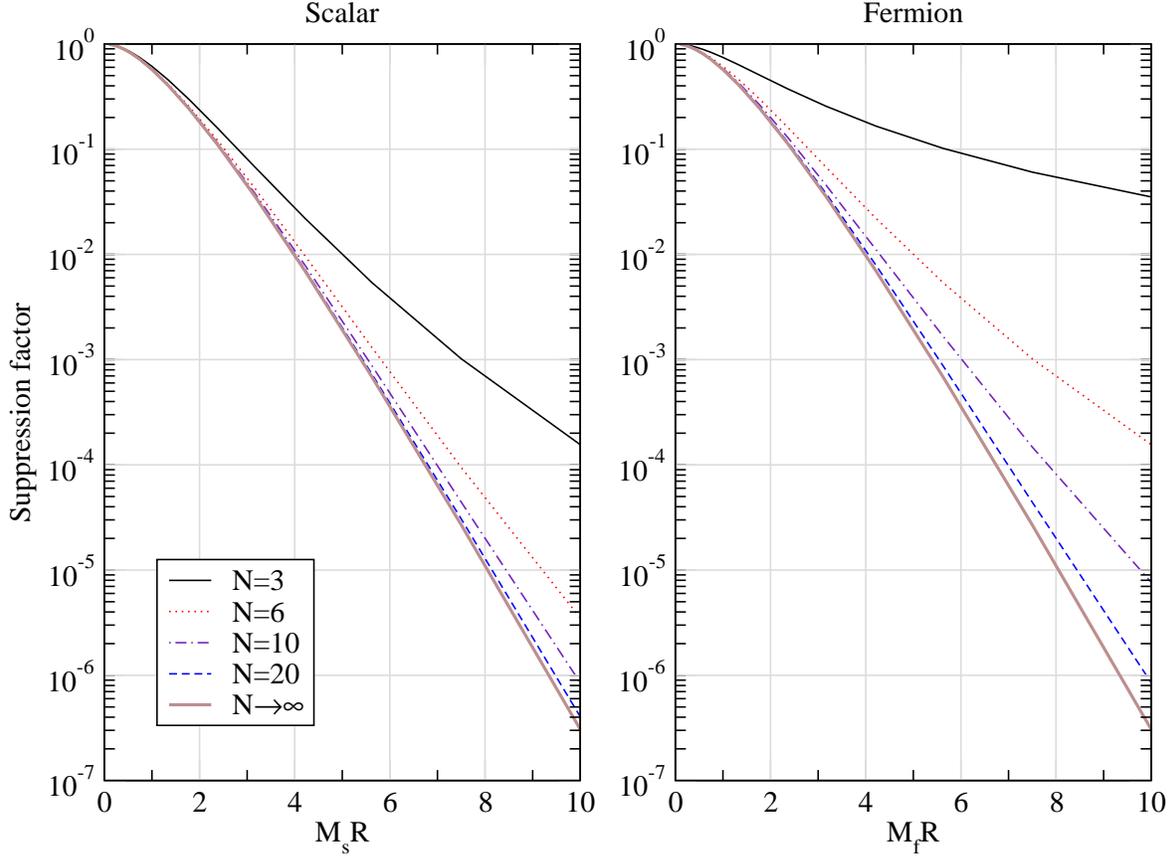}\end{center}

\caption{\label{cap:OffsetMassRatios}The ratios in Eqs.~(\ref{eq:OffsetMassRatioScalar})
and~(\ref{eq:OffsetMassRatioFermion}) of the Casimir energy density
between massive and massless fields. The values for~$N\rightarrow\infty$
are taken from the analytic formulas in Eqs.~(\ref{eq:Rho4ofMUntwistAnalyt})
and~(\ref{eq:Rho4ofMTwistAnalyt}).}
\end{figure}

So far, we have given results only in the case of vanishing kink mass~($M_{\textrm{s,f}}=0$).
For massive five dimensional fields we observe an approximately exponential
suppression of the Casimir energy. This behavior becomes obvious in
the analytical calculation for a continuous extra dimension, which
is given in Sec.~\ref{sec:CasimirAnalytic}. But it is also achieved
for the latticized case of Sec.~\ref{sec:ZeroPointCC}, where in
the limit of an infinite number of lattice sites~($N\rightarrow\infty$)
we approach the values of the analytical formulas~(\ref{eq:Rho4ofMUntwistAnalyt})
and~(\ref{eq:Rho4ofMTwistAnalyt}). To investigate the suppression
behavior depending on the mass~$M_{\textrm{s},\textrm{f}}$ and the
number~$N$, we examine the ratio between the energy density of fields
with mass~$M_{\textrm{s,f}}$ and that of massless fields. For scalar
fields this ratio is defined by\begin{equation}
\frac{S_{1}(M_{\textrm{s}}R)+S_{2}(M_{\textrm{s}}R)}{S_{1}(0)+S_{2}(0)},\label{eq:OffsetMassRatioScalar}\end{equation}
where the functions~$S_{1,2}(N)$ are taken from Eq.~(\ref{eq:CasimirFuncS1})
of Sec.~\ref{sub:CasimirScalar}. Analogously, using the functions~$F_{1,2}(N)$
from Eq.~(\ref{eq:CasimirFuncF1}) of Sec.~\ref{sub:CasimirFermion},
the ratio for fermionic fields reads \begin{equation}
\frac{F_{1}(M_{\textrm{f}}R)+F_{2}(M_{\textrm{f}}R)}{F_{1}(0)+F_{2}(0)}.\label{eq:OffsetMassRatioFermion}\end{equation}
Both ratios are plotted in Fig.~\ref{cap:OffsetMassRatios} for a
range of values of~$N$ and~$M_{\textrm{s},\textrm{f}}R$. The suppression
by a kink mass is most minimal for small~$N$. In the case of~$N=3$
lattice sites, the corresponding ratios are given in Table~\ref{tab:OffsetMassRatioNeq3}.%

\begin{table}[p]
\begin{center}\begin{tabular}{|c|c|c|c|c|c|c|}
\hline 
$M_{\textrm{s,f}}R$&
$1$&
$10$&
$100$&
$1000$&
$10^{4}$&
$10^{5}$\tabularnewline
\hline
\hline 
Scalar&
$0,61$&
$1,5\cdot10^{-4}$&
$2,9\cdot10^{-12}$&
$3,0\cdot10^{-20}$&
$3,0\cdot10^{-28}$&
$3,0\cdot10^{-36}$\tabularnewline
\hline 
Fermion&
$0,74$&
$3,5\cdot10^{-2}$&
$3,7\cdot10^{-4}$&
$3,8\cdot10^{-6}$&
$3,7\cdot10^{-8}$&
$3,7\cdot10^{-10}$\tabularnewline
\hline
\end{tabular}\end{center}

\caption{\label{tab:OffsetMassRatioNeq3}The exponential suppression factors
in Eqs.~(\ref{eq:OffsetMassRatioScalar}) and~(\ref{eq:OffsetMassRatioFermion})
of the Casimir energy densities for $N=3$ lattice sites, where $M_{\textrm{s},\textrm{f}}$
denotes the kink masses of the quantum fields.}
\end{table}

\begin{table}[p]
\begin{center}\begin{tabular}{|c|c|c|c|}
\hline 
field&
inverse lattice spacing $a^{-1}$&
kink mass $M$&
$MR$\tabularnewline
\hline
\hline 
scalars $\Phi_{i}$&
$M_{\textrm{b}}\sim10^{4}-10^{5}\,\textrm{eV}$&
$m\sim10^{2}\,\textrm{GeV}$&
$N\cdot10^{6}$\tabularnewline
\hline 
gauge bosons $A_{i}$&
$u\sim10^{-2}\,\textrm{eV}$&
$v\sim10^{2}\,\textrm{GeV}$&
$N\cdot10^{13}$\tabularnewline
\hline 
fermions $\Psi_{i}$&
$u\sim10^{-2}\,\textrm{eV}$&
$0$&
$0$\tabularnewline
\hline
\end{tabular}\end{center}

\caption{\label{tab:OffsetMassSuppInModel}The fields which can be interpreted
as KK modes of a 5D field on a transverse lattice. The kink mass is
denoted by $M$, and $R=Na$ is the circumference of $\mathcal{S}_{\textrm{lat}}^{1}$
with $N$ sites and a lattice spacing $a$. For large values of~$MR$
the Casimir effect will be highly suppressed, see Fig.~\ref{cap:OffsetMassRatios}
and Table~\ref{tab:OffsetMassRatioNeq3}.}
\end{table}

%\newpage

\section{Zero-point Energy in Deconstruction\label{sec:ZeroPointDeconst}}

We will now determine the finite value to the formally infinite vacuum
energies of the 4D quantum fields in deconstruction. In the non-linear
sigma model approximation, deconstruction yields a transverse lattice
description%
\footnote{The transverse lattice can, in general, respect perturbative unitarity
up to energies~$\sqrt{s}\gtrsim2\pi N/R$~\cite{Chivukula:2002ej}.%
} of the~$\mathcal{M}\times\mathcal{S}_{\textrm{lat}}^{1}$ space-time
for which we have already calculated the Casimir energies of different
fields species in Sec.~\ref{sec:ZeroPointCC}. Then, by correspondence
with the parent 5D theory, we are in a position to adopt the renormalization
procedure from the Casimir effect of a 5D quantum field to renormalize
the divergent vacuum energies of the associated 4D effective KK modes
in the deconstruction setup. Specifically, in the model of Sec.~\ref{sec:Model},
all fields, except the fermionic ones, have large kink masses. As
a result of Table~\ref{tab:OffsetMassSuppInModel} and Sec.~\ref{sub:OffsetMass},
the Casimir effect is therefore highly suppressed only for the bosons.
Consequently, the fermionic fields~$\Psi_{i}$~($i=1,\dots,N$)
provide the dominant part to~$\Lambda$, and their total vacuum energy
density is given by the sum of zero-point energies\begin{equation}
\rho=-\frac{2}{(2\pi)^{3}}\sum_{n=1}^{N}\int\textrm{d}^{3}p\,\omega_{n}(p),\label{eq:ZeroPoint4DFermions}\end{equation}
where the $n$th field has a mass~$M_{n}$ and momentum modes with
energy\[
\omega_{n}(p)=\sqrt{p^{2}+M_{n}^{2}}.\]
We will now show that the masses~$M_{n}$ follow exactly the momentum
spectrum of the 5D fermion field in Eq.~(\ref{eq:FermionEngMomRel}),
implying that the expression in Eq.~(\ref{eq:ZeroPoint4DFermions})
becomes identical with the effective unrenormalized Casimir energy
density obtained from Eq.~(\ref{eq:FermionRhoUnregul}). This clearly
reflects also on a formal level the correspondence between the deconstructed
and the 5D theory, by which we can apply the Casimir renormalization
techniques of Sec.~\ref{sec:ZeroPointCC} to the energy density $\rho$
in Eq.~(\ref{eq:ZeroPoint4DFermions}). To this end, we consider
first the kinetic term in Eq.~(\ref{eq:typeIImasses}). When the
link fields~$Q_{i}$ acquire universal VEVs~$u$ after SSB,~$\mathcal{L}_{\textrm{mass}}[\Psi_{i},Q_{i}]$
takes the form \[
\mathcal{L}_{\textrm{mass}}[\Psi_{i},Q_{i}]\rightarrow\frac{u}{2}\cdot\sum_{n=1}^{N}\overline{\Psi}_{n\textrm{L}}\left[\Psi_{(n+1)\textrm{R}}-\Psi_{(n-1)\textrm{R}}\right]+\textrm{h.c.},\]
where~$\Psi_{n\textrm{L,R}}=\frac{1}{2}(1\mp\gamma_{5})\Psi_{n}$
are the left- and right-handed chiral components of the Dirac spinor~$\Psi_{n}=(\Psi_{n\textrm{L}},\Psi_{n\textrm{R}})^{\textrm{T}}$.
The sum in~$\mathcal{L}_{\textrm{mass}}[\Psi_{i},Q_{i}]$ contains
{}``boundary terms''~$-\frac{u}{2}\overline{\Psi}_{1\textrm{L}}\Psi_{0\textrm{R}}$
and~$\frac{u}{2}\overline{\Psi}_{N\textrm{L}}\Psi_{(N+1)\textrm{R}}$,
where~$\Psi_{0\textrm{R}}$ and~$\Psi_{(N+1)\textrm{R}}$ are defined
by\[
\Psi_{mN+1}=T^{m}\cdot\Psi_{1}\,\,\,\,\textrm{and}\,\,\,\,\Psi_{0}=T^{m}\cdot\Psi_{mN},\,\,\,\, m\in\mathbb{Z},\]
where we distinguish between untwisted~($T=+1$) and twisted~($T=-1$)
fermionic fields. Note that this is the discretized version of the
continuum boundary condition\[
\Psi(y+mR)=T^{m}\cdot\Psi(y),\,\,\,\, m\in\mathbb{Z}.\]
 Now, the Lagrangian in matrix form reads\[
\mathcal{L}_{\textrm{mass}}[\Psi_{i}]=\sum_{n,k=1}^{N}\overline{\Psi}_{n\textrm{L}}M_{nk}\Psi_{k\textrm{R}}+\textrm{h.c.},\]
where the mass matrix~$M_{nk}$ and its square~$M^{2}=MM^{\dag}=M^{\dag}M$
are explicitly given by\[
M=\frac{u}{2}\left[\begin{array}{rrrcr}
0 & +1 &  &  & -T\\
-1 & 0 & +1\\
\\ & \ddots & \ddots & \ddots\\
\\ &  & -1 & 0 & +1\\
+T &  &  & -1 & 0\end{array}\right],\,\,\,\, M^{2}=\frac{u^{2}}{4}\left[\begin{array}{rrrrrrr}
2 & 0 & -1 &  &  & -T & 0\\
0 & 2 & 0 & -1 &  &  & -T\\
-1 & 0 & 2 & 0 & -1\\
 & \ddots & \ddots & \ddots & \ddots & \ddots\\
 &  & -1 & 0 & 2 & 0 & -1\\
-T &  &  & -1 & 0 & 2 & 0\\
0 & -T &  &  & -1 & 0 & 2\end{array}\right].\]
The squared masses~$m_{n}^{2}$ of the fermions are found to be the
eigenvalues of~$M^{2}$. Thus, the mass spectrum for untwisted fields
reads\[
m_{n}^{2}=u^{2}\sin^{2}2\pi\frac{n}{N},\,\,\,\, n=1,\dots,N,\]
and for the twisted fields we obtain\[
m_{n}^{2}=u^{2}\sin^{2}2\pi\frac{n-\frac{1}{2}}{N},\,\,\,\, n=1,\dots,N,\]
which is consistent with the results in Ref.~\cite{Hill:2002me}.
Note that only for odd~$N$, both spectra become identical, which
has already been discussed in the context of the odd-even artefact
in Sec.~\ref{sub:CasimirSummary}. With these mass spectra the Casimir
energy of the fermions is~$(-8)$ times the Casimir energy of a real
scalar, whereas in a non-lattice calculation the energies differ only
by a factor of~$(-4)$ (see Sec.~\ref{sub:CasimirSummary}). This
comes again from the well known phenomenon of fermion doubling in
lattice theory. One way to remove this problem is to add a \emph{Wilson
term~}\cite{Wilson:1974sk} to~$\mathcal{L}_{\textrm{mass}}[\Psi_{i},Q_{i}]$
in Eq.~(\ref{eq:typeIImasses}). This then modifies the Lagrangian
in Eq.~(\ref{eq:typeIImasses}):\begin{equation}
\mathcal{L}_{\textrm{mass}}=u\cdot\sum_{n=1}^{N}\left[\overline{\Psi}_{n\textrm{L}}\left(\frac{Q_{n}^{\dagger}}{u}\Psi_{(n+1)\textrm{R}}-\Psi_{n\textrm{R}}\right)-\overline{\Psi}_{n\textrm{R}}\left(\Psi_{n\textrm{L}}-\frac{Q_{n-1}}{u}\Psi_{(n-1)\textrm{L}}\right)\right],\label{eq:WilsonMassLagr}\end{equation}
which, after SSB, yields the mass matrix~$M$. The squared masses~$m_{n}^{2}$
of the fermions are now the eigenvalues of~$M^{2}$:\[
M=u\cdot\left[\begin{array}{rrrrr}
-1 & +1 &  &  & 0\\
 & -1 & +1\\
 &  & \ddots & \ddots\\
 &  &  & -1 & +1\\
+T &  &  &  & -1\end{array}\right],\,\,\,\, M^{2}=u^{2}\cdot\left[\begin{array}{rrrrr}
2 & -1 &  &  & -T\\
-1 & 2 & -1\\
 & \ddots & \ddots & \ddots\\
 &  & -1 & 2 & -1\\
-T &  &  & -1 & 2\end{array}\right],\]
For the untwisted fields we get\begin{equation}
m_{n}^{2}=2u^{-2}\left(1-\cos2\pi\frac{n}{N}\right),\,\,\,\, n=1\dots N,\label{eq:MassSpecScalarWilsFerm}\end{equation}
and for the twisted ones~$n$ is replaced by~$n-\frac{1}{2}$. These
mass spectra are identical with the mass spectra of real scalars and
yields the usual (continuum) factor~$(-4)$ in the vacuum energy
density between the two field species. The Wilson term successfully
prevented the fermion doubling. Looking at the above mass spectra,
we remark that the spectrum~(\ref{eq:MassSpecScalarWilsFerm}) for
scalars and Wilson-modified fermions does not contain a zero mode
in the case of twisted fields.

The Wilson term in Eq.~(\ref{eq:WilsonMassLagr}) involves explicit
Dirac mass terms of the type~$u\cdot\overline{\Psi}_{n\textrm{L}}\Psi_{n\textrm{R}}$.
Let us now examine a possibility to generate such small Dirac masses~$u\sim10^{-2}\,\textrm{eV}$
in a natural way. For this purpose, we assume two extra SM singlet
scalars~$Q_{0}$ and~$\Phi_{0}$, where~$Q_{0}$ has the large
mass~$M\sim10^{8}\,\textrm{GeV}$ and~$\Phi_{0}$ has the mass~$m\sim10^{2}\,\textrm{GeV}$.
Also,~$Q_{0}$ and~$\Phi_{0}$ are singlets under the product group~$U(1)^{N}$.
Next, we suppose a discrete~$Z_{2N}$ symmetry acting on the fields
as follows:\begin{eqnarray*}
 & Z_{2N}: & \Phi_{n}\rightarrow e^{i2\pi n/N}\cdot\Phi_{n},\,\,\,\, Q_{n}\rightarrow e^{i2\pi/N}\cdot Q_{n},\,\,\,\,\Psi_{n\textrm{L}}\rightarrow e^{-i2\pi/N}\cdot\Psi_{n\textrm{L}},\\
 &  & \Phi_{0}\rightarrow e^{i\pi/N}\cdot\Phi_{0},\,\,\,\, Q_{0}\rightarrow e^{i2\pi/N}\cdot Q_{0},\,\,\,\,\,\,\,\, n=1,\dots,N\end{eqnarray*}
 All Yukawa interactions of the type~$\overline{\Psi}_{n\textrm{L}}Q_{n}^{\dag}\Psi_{(n+1)\textrm{R}}/u$
are left unaffected by this symmetry in the Wilson term which forbids
all explicit mass terms. The only Yukawa interaction which is, in
addition, allowed by the $Z_{2N}$-symmetry is of the form~$\overline{\Psi}_{n\textrm{L}}Q_{0}^{\dag}\Psi_{n\textrm{R}}$,
coupling left- and right-handed states sitting on the same lattice
site. Schematically, in presence of the fields~$Q_{0}$ and~$\Phi_{0}$,
the potential in Eq.~(\ref{eq:seesawpotential}) is modified as follows:\begin{eqnarray}
V & \rightarrow & V+m^{2}\Phi_{0}^{\dag}\Phi_{0}+M^{2}Q_{0}^{\dag}Q_{0}+\frac{1}{2}\lambda_{1}(\Phi_{0}^{\dag}\Phi_{0})^{2}+\frac{1}{2}\lambda_{2}(Q_{0}^{\dag}Q_{0})^{2}\nonumber \\
 &  & +(\lambda_{3}Q_{0}^{\dag}Q_{0}+\lambda_{4}\Phi_{0}^{\dag}\Phi_{0})\sum_{j=1}^{N}\Phi_{j}^{\dag}\Phi_{j}+\mu\Phi_{0}^{\dag}Q_{0}\Phi_{0}^{\dag}+\mu^{*}\Phi_{0}Q_{0}^{\dag}\Phi_{0}\nonumber \\
 &  & +(\lambda_{3}\Phi_{0}^{\dag}\Phi_{0}+\lambda_{4}Q_{0}^{\dag}Q_{0})\sum_{j=1}^{N}Q_{j}^{\dag}Q_{j},\label{eq:ModSeesawPot}\end{eqnarray}
where~$\lambda_{1},\dots,\lambda_{4}$ are order unity coefficients.
It is thus seen, that the interactions of the scalars~$Q_{0}$ and~$\Phi_{0}$
also reproduce the structure of~$V$ which is relevant for the type-II
seesaw mechanism. We therefore expect the mechanism in Sec.~\ref{sec:Model}
for generating \mbox{sub-mm} lattice spacings to be valid (at least
qualitatively) also for the modified potential in Eq.~(\ref{eq:ModSeesawPot}).
From this we find, that the choice~$m^{2}<0$ and~$M^{2}>0$ leads
to VEVs of the orders~$\left\langle \Phi_{0}\right\rangle \sim10^{2}\,\textrm{GeV}$
and~$\left\langle Q_{0}\right\rangle \sim10^{-2}\,\textrm{eV}$.
We hence conclude, that a small Dirac mass~$u$ for the mass terms~$u\cdot\overline{\Psi}_{n\textrm{L}}\Psi_{n\textrm{R}}$
in Eq.~(\ref{eq:WilsonMassLagr}) can be generated spontaneously
from the Yukawa interaction~$\overline{\Psi}_{n\textrm{L}}Q_{0}^{\dag}\Psi_{nR}$
in a natural way thereby providing an understanding of the mass scales
involved in the Wilson term.

The discussion of mass spectra can be easily repeated for the bosonic
fields in the model as well. Therefore, we identify the physical vacuum
energies of the fields in the model of dimensional deconstruction
of Sec.~\ref{sec:Model} with the finite Casimir results of the calculation
in Sec.~\ref{sec:ZeroPointCC}. In Table~\ref{tab:OffsetMassSuppInModel}
we give the relevant quantities for the vacuum energy suppression
of the fields. Consequently, the Casimir energies of the scalar and
vector fields can be neglected due to their large kink masses of order~$10^{2}\,\textrm{GeV}$.
In contrast to this, the fermionic fields with KK masses of the order
of the small VEV~$u\sim10^{-2}\,\textrm{eV}$ from Eq.~(\ref{eq:SmallVEVu})
induce a positive contribution to~$\Lambda$ which is of the observed
order of magnitude already for a small number of~$N\lesssim10$ lattice
sites. Finally, it should be noted that we determined only the vacuum
energy contributions of quantum fields to~$\Lambda$, not its absolute
value.

%\newpage

\section{The Casimir Effect in the Continuum\label{sec:CasimirAnalytic}}

In this section, the Casimir effect for a continuous extra dimension
will be investigated analytically, where we put special emphasis on
the dependence on the mass of the quantum field. In the continuum
case, there is no difference in the momentum sums/integrals between
fermions and scalars, in both cases the energy function~$\omega$
has the form \[
\omega^{2}=\vec{p}^{\,2}+q^{2}+M_{\textrm{s,f}}^{2},\]
where~$q=2\pi n/R$ and~$q=2\pi(n+\frac{1}{2})/R$ for untwisted
and twisted fields, respectively. We also have to consider negative
and unbounded momenta~$q$ in the mode sum. So the unrenormalized
energy density is\[
\rho\propto\int\textrm{d}^{3}p\sum_{n=-\infty}^{\infty}\omega.\]
After the regularization used in Eq.~(\ref{eq:ScalarRhoRegul}) of
the~$\int\textrm{d}^{3}p$-integral, the remaining sum over the discrete
five-momentum~$q=2\pi n/R$ can be calculated by using the Abel-Plana
formulas~\cite{Saharian:2000xx} as a renormalization prescription:\begin{eqnarray}
\sum_{n=0}^{\infty}f(n)-\int_{0}^{\infty}\textrm{d}n\cdot f(n) & = & \frac{1}{2}f(0)+i\int_{0}^{\infty}\textrm{d}n\cdot\frac{f(+in)-f(-in)}{\exp(2\pi n)-1},\label{eq:AbelPlana1}\\
\sum_{n=0}^{\infty}f(n+\kl{\frac{1}{2}})-\int_{0}^{\infty}\textrm{d}n\cdot f(n) & = & -i\int_{0}^{\infty}\textrm{d}n\cdot\frac{f(+in)-f(-in)}{\exp(2\pi n)+1}.\label{eq:AbelPlana2}\end{eqnarray}
The subtraction of the integral on the left hand side corresponds
to the subtraction in the first line of Eq.~(\ref{eq:ScalarRho5Ren}).
In the case of untwisted fields we have\[
A:=\sum_{n=-\infty}^{\infty}m^{4}(n)\ln[m(n)]-\int_{-\infty}^{\infty}\textrm{d}n\cdot m^{4}(n)\ln[m(n)],\]
where~$m(n)=\sqrt{k^{2}n^{2}+M^{2}}$ and~$k:=2\pi/R$, and with
the first Abel-Plana formula~(\ref{eq:AbelPlana1}) we rewrite this
difference:\[
A=2i\int_{0}^{\infty}\textrm{d}n\frac{m^{4}(+in)\ln[m(+in)]-m^{4}(-in)\ln[m(-in)]}{\exp(2\pi n)-1}.\]
Assuming~$k,n,M\geq0$, we must consider two cases for the root~$m(\pm in)$:\[
\left[k^{2}(\pm in)^{2}+M^{2}\right]^{\frac{1}{2}}=\left\{ \begin{array}{ll}
(M^{2}-k^{2}n^{2})^{\frac{1}{2}} & \,\,\,\textrm{for}\,\,\, M>kn,\\
\pm i(k^{2}n^{2}-M^{2})^{\frac{1}{2}} & \,\,\,\textrm{for}\,\,\, M<kn.\end{array}\right.\]
For~$a>0$ we can write the logarithm as $\ln(\pm i\cdot a)=\pm i\pi/2+\ln a$,
and with~$x:=M/k$ we obtain the result\begin{equation}
A(M)=-2\pi k^{4}\int_{x}^{\infty}\textrm{d}n\frac{(n^{2}-x^{2})^{2}}{\exp(2\pi n)-1},\label{eq:CasAnalytAofM}\end{equation}
which has, in the massless case~($x=0$), the value \[
A=-2\pi k^{4}\frac{3}{4\pi^{5}}\cdot\zeta(5)=-8R^{-4}3\zeta(5).\]

For twisted fields we have to use the second Abel-Plana formula~(\ref{eq:AbelPlana2}).
Analogously, we write\[
B(M)=-2i\int_{0}^{\infty}\textrm{d}n\frac{m^{4}(+in)\ln[m(+in)]-m^{4}(-in)\ln[m(-in)]}{\exp(2\pi n)+1},\]
using~$m(n)=\sqrt{k^{2}n^{2}+M^{2}}$, and not~$m(n)=\sqrt{k^{2}(n+\frac{1}{2})^{2}+M^{2}}$.
Thus, the result becomes\begin{equation}
B(M)=+2\pi k^{4}\int_{x}^{\infty}\textrm{d}n\frac{(n^{2}-x^{2})^{2}}{\exp(2\pi n)+1},\label{eq:CasAnalytBofM}\end{equation}
where the value for massless fields~($x=0$) is\[
B=+2\pi k^{4}\frac{45}{64\pi^{5}}\cdot\zeta(5)=\frac{15}{2}R^{-4}3\zeta(5).\]
For large masses~($x\gg1$), approximate expressions for~$A(M)$
and~$B(M)$ can be given by neglecting the~$1$ in the denominator
of the Eqs.~(\ref{eq:CasAnalytAofM}) and~(\ref{eq:CasAnalytBofM}):\[
\int_{x}^{\infty}\textrm{d}n\frac{(n^{2}-x^{2})^{2}}{\exp(2\pi n)\pm1}\stackrel{x\gg1}{\sim}\int_{x}^{\infty}\textrm{d}n\frac{(n^{2}-x^{2})^{2}}{\exp(2\pi n)}=\frac{4\pi^{2}x^{2}+6\pi x+3}{4\pi^{5}}e^{-2\pi x}.\]
This shows the (approximately) exponential suppression of the Casimir
energy by the field mass~$M=xk$. Using Eqs.~(\ref{eq:ScalarRho5Ren})
and~(\ref{eq:CasAnalytAofM}) the effective energy density~$\rho_{4}=R\rho$
for untwisted scalar fields is given by\begin{eqnarray}
\rho_{4} & = & \frac{1}{8(2\pi)^{2}}\cdot A(M)\,\,\stackrel{M=0}{=}\,\,-1\cdot(2\pi)^{-2}R^{-4}3\zeta(5),\label{eq:Rho4ofMUntwistAnalyt}\end{eqnarray}
and with Eq.~(\ref{eq:CasAnalytBofM}) we obtain the result for the
twisted fields:\begin{eqnarray}
\rho_{4} & = & \frac{1}{8(2\pi)^{2}}\cdot B(M)\,\,\stackrel{M=0}{=}\,\,+\frac{15}{16}\cdot(2\pi)^{-2}R^{-4}3\zeta(5).\label{eq:Rho4ofMTwistAnalyt}\end{eqnarray}
The energy densities for Dirac fermions are just~$(-4)$ times the
values for the real scalars.

%\newpage

\section{Summary and Conclusions\label{sec:SummaryConclusions}}

In this paper, we have shown a way how to calculate finite zero-point
energies of 4D quantum fields which have a higher-dimensional correspondence
in deconstruction. In particular, we constructed a 4D~$U(1)^{N}$
model which mimics a latticized large extra dimension with lattice
spacings~$a$ in the sub-mm range. The vacuum energy of the fermions
in our model gives rise to a small and positive contribution~$\rho\sim(Na)^{-4}\sim(10^{-3}\,\textrm{eV})^{4}$
to the cosmological constant in agreement with recent observations.
The negative contributions of the scalars and gauge bosons, on the
other hand, are exponentially suppressed by large kink masses. Here,
we used the correspondence between the 4D zero-point energies and
the unrenormalized Casimir energy of 5D quantum fields in a geometric
transverse lattice space-time. With inverse lattice spacings in the
sub-eV range, our mechanism allows to dynamically generate a large
compactified extra dimension with only a small number of lattice sites.
This is achieved by giving the link fields a large mass of the order~$10^{8}\,\textrm{GeV}$
and a bulk scalar with kink mass in the electroweak range. Alternatively,
this scalar can also be interpreted as a link variable in a space
which is topologically a disk. The Casimir effect on the transverse
lattice has been investigated for scalar and fermion fields in more
detail, thereby taking into account twisted and untwisted field configurations
which arise in multiply connected space-times. For the fermions, we
observed an odd-even artefact in the Casimir energy which disappears
when taking the energy sum of twisted and untwisted fermionic fields.
Moreover, due to the naive discretization procedure, we also encountered
the effect of fermion doubling, which has been removed by the usual
Wilson modification of the fermionic kinetic terms. Furthermore, the
suppression of the Casimir effect by a kink mass has been shown for
fields on the lattice and in the continuum. Although, this has been
utilized only to neglect unwanted contributions to the vacuum energy,
it could also be used to generate tiny energy values for quantum fields
in small extra dimensions. The combination of methods and mechanisms
employed in this work may be generalized for other purposes, and a
deeper discussion of vacuum energy in non-trivial topologies poses
a task in the future.

\section*{Acknowledgments}

We would like to thank T.~Ohlsson and S.~Pokorski for useful comments
and discussions. This work was supported by the {}``Sonderforschungsbereich
375 für Astroteilchenphysik der Deutschen Forschungsgemeinschaft''.
F.B. wishes to thank the Freistaat Bayern for financial support by
a {}``Landesgraduiertenstipendium''.

\appendix

\section{Minimization of the Potential\label{secA:MinimizPotential}}

We will minimize the scalar potential~$V$ in Eq.~(\ref{eq:seesawpotential})
by going to the real basis in Eq.~(\ref{eq:realBasis}). In this
parameterization, the term $\mu\Phi_{i-1}Q_{i}\Phi_{i}^{\dagger}$
in Eq.~(\ref{eq:seesawpotential}) reads \begin{eqnarray}
\mu\Phi_{i-1}Q_{i}\Phi_{i}^{\dagger} & = & \mu\left[\phi_{i-1}^{a}q_{i}^{a}\phi_{i}^{a}-\phi_{i-1}^{b}q_{i}^{b}\phi_{i}^{a}+\phi_{i-1}^{b}q_{i}^{a}\phi_{i}^{b}+\phi_{i-1}^{a}q_{i}^{b}\phi_{i}^{b}\right.\nonumber \\
 &  & \left.+{\textrm{i}}(\phi_{i-1}^{b}q_{i}^{a}\phi_{i}^{a}+\phi_{i-1}^{a}q_{i}^{b}\phi_{i}^{a}-\phi_{i-1}^{a}q_{i}^{a}\phi_{i}^{b}+\phi_{i-1}^{b}q_{i}^{b}\phi_{i}^{b})\right].\label{eq:muterm}\end{eqnarray}
 Also, the term $\lambda_{6}(Q_{i}Q_{i+1})(\Phi_{i}\Phi_{i+2}^{\dagger})$
in Eq.~(\ref{eq:seesawpotential}) is given by \begin{eqnarray}
\lambda_{6}(Q_{i}Q_{i+1})(\Phi_{i}\Phi_{i+2}^{\dagger}) & = & \lambda_{6}\left[(q_{i}^{a}q_{i+1}^{a}-q_{i}^{b}q_{i+1}^{b})(\phi_{i}^{a}\phi_{i+2}^{a}+\phi_{i}^{b}\phi_{i+2}^{b})\right.\nonumber \\
 &  & \left.-(q_{i}^{b}q_{i+1}^{a}+q_{i}^{a}q_{i+1}^{b})(\phi_{i}^{b}\phi_{i+2}^{a}-\phi_{i}^{a}\phi_{i+2}^{b})\right]\nonumber \\
 &  & +{\textrm{i}}\lambda_{6}\left[(q_{i}^{a}q_{i+1}^{a}-q_{i}^{b}q_{i+1}^{b})(\phi_{i}^{b}\phi_{i+2}^{a}-\phi_{i}^{a}\phi_{i+2}^{b})\right.\nonumber \\
 &  & \left.+(q_{i}^{b}q_{i+1}^{a}+q_{i}^{a}q_{i+1}^{b})(\phi_{i}^{a}\phi_{i+2}^{a}+\phi_{i}^{b}\phi_{i+2}^{b})\right].\label{eq:lambda6term}\end{eqnarray}

Then, the scalar potential in Eq.~(\ref{eq:seesawpotential}) can
be written as \begin{eqnarray*}
V & = & \sum_{i=1}^{N}\left\{ m^{2}\left[(\phi_{i}^{a})^{2}+(\phi_{i}^{b})^{2}\right]+M^{2}\left[(q_{i}^{a})^{2}+(q_{i}^{b})^{2}\right]+\frac{1}{2}\lambda_{1}\left[(\phi_{i}^{a})^{2}+(\phi_{i}^{b})^{2}\right]^{2}\right.\\
 &  & +\frac{1}{2}\lambda_{2}\left[(q_{i}^{a})^{2}+(q_{i}^{b})^{2}\right]^{2}+\lambda_{3}\left[(\phi_{i}^{a})^{2}+(\phi_{i}^{b})^{2}\right]\left[\sum_{j=1}^{N}(q_{j}^{a})^{2}+(q_{j}^{b})^{2}\right]\\
 &  & +\lambda_{4}\left[(\phi_{i}^{a})^{2}+(\phi_{i}^{b})^{2}\right]\left[\sum_{j\neq i}(\phi_{j}^{a})^{2}+(\phi_{j}^{b})^{2}\right]\\
 &  & +\lambda_{5}\left[(q_{i}^{a})^{2}+(q_{i}^{b})^{2}\right]\left[\sum_{j\neq i}(q_{j}^{a})^{2}+(q_{j}^{b})^{2}\right]\\
 &  & +\mu\left[\phi_{i-1}^{a}q_{i}^{a}\phi_{i}^{a}-\phi_{i-1}^{b}q_{i}^{b}\phi_{i}^{a}+\phi_{i-1}^{b}q_{i}^{a}\phi_{i}^{b}+\phi_{i-1}^{a}q_{i}^{b}\phi_{i}^{b}\right]\\
 &  & +\mu\left[\phi_{i}^{a}q_{i+1}^{a}\phi_{i+1}^{a}-\phi_{i}^{b}q_{i+1}^{b}\phi_{i+1}^{a}+\phi_{i}^{b}q_{i+1}^{a}\phi_{i+1}^{b}+\phi_{i}^{a}q_{i+1}^{b}\phi_{i+1}^{b}\right]\\
 &  & +\frac{1}{2}\lambda_{6}\left[(q_{i-2}^{a}q_{i-1}^{a}-q_{i-2}^{b}q_{i-1}^{b})(\phi_{i-2}^{a}\phi_{i}^{a}+\phi_{i-2}^{b}\phi_{i}^{b})\right.\\
 &  & \left.-(q_{i-2}^{b}q_{i-1}^{a}+q_{i-2}^{a}q_{i-1}^{b})(\phi_{i-2}^{b}\phi_{i}^{a}-\phi_{i-2}^{a}\phi_{i}^{b})\right]\\
 &  & +\frac{1}{2}\lambda_{6}\left[(q_{i-1}^{a}q_{i}^{a}-q_{i-1}^{b}q_{i}^{b})(\phi_{i-1}^{a}\phi_{i+1}^{a}+\phi_{i-1}^{b}\phi_{i+1}^{b})\right.\\
 &  & \left.-(q_{i-1}^{b}q_{i}^{a}+q_{i-1}^{a}q_{i}^{b})(\phi_{i-1}^{b}\phi_{i+1}^{a}-\phi_{i-1}^{a}\phi_{i+1}^{b})\right]\end{eqnarray*}
 \begin{eqnarray}
 &  & +\frac{1}{2}\lambda_{6}\left[(q_{i}^{a}q_{i+1}^{a}-q_{i}^{b}q_{i+1}^{b})(\phi_{i}^{a}\phi_{i+2}^{a}+\phi_{i}^{b}\phi_{i+2}^{b})\right.\nonumber \\
 &  & \left.-(q_{i}^{b}q_{i+1}^{a}+q_{i}^{a}q_{i+1}^{b})(\phi_{i}^{b}\phi_{i+2}^{a}-\phi_{i}^{a}\phi_{i+2}^{b})\right]\nonumber \\
 &  & +\frac{1}{2}\lambda_{6}\left[(q_{i+1}^{a}q_{i+2}^{a}-q_{i+1}^{b}q_{i+2}^{b})(\phi_{i+1}^{a}\phi_{i+3}^{a}+\phi_{i+1}^{b}\phi_{i+3}^{b})\right.\nonumber \\
 &  & \left.\left.-(q_{i+1}^{b}q_{i+2}^{a}+q_{i+1}^{a}q_{i+2}^{b})(\phi_{i+1}^{b}\phi_{i+3}^{a}-\phi_{i+1}^{a}\phi_{i+3}^{b})\right]\right\} ,\label{eq:explicitpotential}\end{eqnarray}
where we have symmetrically reorganized the sum, such that all operators
carrying the index {}``$i$'' are explicitly displayed%
\footnote{To avoid double-counting, the coefficients $\mu$ and $\lambda_{6}$
have been given pre-factors $\frac{1}{2}$ and $\frac{1}{4}$, respectively.%
}. We are interested in a minimum of $V$ with a vacuum structure as
given in Eq.~(\ref{eq:universalVEVs}), \textit{i.e.}, all link variables
$Q_{i}$ have a real universal VEV $u$ and all fields~$\Phi_{i}$
have a real universal VEV $v$. From Eq.~(\ref{eq:explicitpotential})
we obtain \begin{eqnarray}
\frac{\partial V}{\partial\phi_{i}^{a}} & = & 2m^{2}\phi_{i}^{a}+2\lambda_{1}\left[(\phi_{i}^{a})^{2}+(\phi_{i}^{b})^{2}\right]\phi_{i}^{a}\nonumber \\
 &  & +2\lambda_{3}\phi_{i}^{a}\left[\sum_{j=1}^{N}(q_{j}^{a})^{2}+(q_{j}^{b})^{2}\right]+2\lambda_{4}\phi_{i}^{a}\left[\sum_{j\neq i}(\phi_{j}^{a})^{2}+(\phi_{j}^{b})^{2}\right]\nonumber \\
 &  & +\mu[\phi_{i-1}^{a}q_{i}^{a}-\phi_{i-1}^{b}q_{i}^{b}]+\mu\left[q_{i+1}^{a}\phi_{i+1}^{a}+q_{i+1}^{b}\phi_{i+1}^{b}\right]\nonumber \\
 &  & +\frac{1}{2}\lambda_{6}\left[(q_{i-2}^{a}q_{i-1}^{a}-q_{i-2}^{b}q_{i-1}^{b})\phi_{i-2}^{a}-(q_{i-2}^{b}q_{i-1}^{a}+q_{i-2}^{a}q_{i-1}^{b})\phi_{i-2}^{b}\right]\nonumber \\
 &  & +\frac{1}{2}\lambda_{6}\left[(q_{i}^{a}q_{i+1}^{a}-q_{i}^{b}q_{i+1}^{b})\phi_{i+2}^{a}+(q_{i}^{b}q_{i+1}^{a}+q_{i}^{a}q_{i+1}^{b})\phi_{i+2}^{b}\right],\label{eq:derivVphia}\\
\nonumber \\\frac{\partial V}{\partial\phi_{i}^{b}} & = & 2m^{2}\phi_{i}^{b}+2\lambda_{1}\left[(\phi_{i}^{a})^{2}+(\phi_{i}^{b})^{2}\right]\phi_{i}^{b}\nonumber \\
 &  & +2\lambda_{3}\phi_{i}^{b}\left[\sum_{j=1}^{N}(q_{j}^{a})^{2}+(q_{j}^{b})^{2}\right]+2\lambda_{4}\phi_{i}^{b}\left[\sum_{j\neq i}(\phi_{j}^{a})^{2}+(\phi_{j}^{b})^{2}\right]\nonumber \\
 &  & +\mu[\phi_{i-1}^{b}q_{i}^{a}+\phi_{i-1}^{a}q_{i}^{b}]+\mu\left[-q_{i+1}^{b}\phi_{i+1}^{a}+q_{i+1}^{a}\phi_{i+1}^{b}\right]\nonumber \\
 &  & +\frac{1}{2}\lambda_{6}\left[(q_{i-2}^{a}q_{i-1}^{a}-q_{i-2}^{b}q_{i-1}^{b})\phi_{i-2}^{b}+(q_{i-2}^{b}q_{i-1}^{a}+q_{i-2}^{a}q_{i-1}^{b})\phi_{i-2}^{a}\right]\nonumber \\
 &  & +\frac{1}{2}\lambda_{6}\left[(q_{i}^{a}q_{i+1}^{a}-q_{i}^{b}q_{i+1}^{b})\phi_{i+2}^{b}-(q_{i}^{b}q_{i+1}^{a}+q_{i}^{a}q_{i+1}^{b})\phi_{i+2}^{a}\right],\label{eq:derivVphib}\end{eqnarray}
 which gives for the VEVs in Eq.~(\ref{eq:universalVEVs}) the minimization
condition\[
m^{2}+\left[\lambda_{1}+(N-1)\lambda_{4}\right]v^{2}+(N\lambda_{3}+\frac{1}{2}\lambda_{6})u^{2}+\mu u=0,\]
 and $\langle\partial V/\partial\phi_{i}^{b}\rangle=0$ is automatic
for these VEVs. The partial derivatives for the link fields $Q_{i}$
are \begin{eqnarray}
\frac{\partial V}{\partial q_{i}^{a}} & = & 2M^{2}q_{i}^{a}+2\lambda_{2}\left[(q_{i}^{a})^{2}+(q_{i}^{b})^{2}\right]q_{i}^{a}+2\lambda_{3}q_{i}^{a}\left[\sum_{j=1}^{N}(\phi_{j}^{a})^{2}+(\phi_{j}^{b})^{2}\right]\nonumber \\
 &  & +\mu(\phi_{i-1}^{a}\phi_{i}^{a}+\phi_{i-1}^{b}\phi_{i}^{b})+2\lambda_{5}q_{i}^{a}\left[\sum_{j\neq i}(q_{j}^{a})^{2}+(q_{j}^{b})^{2}\right]\nonumber \\
 &  & +\frac{1}{2}\lambda_{6}\left[q_{i-1}^{a}(\phi_{i-1}^{a}\phi_{i+1}^{a}+\phi_{i-1}^{b}\phi_{i+1}^{b})-q_{i-1}^{b}\left(\phi_{i-1}^{b}\phi_{i+1}^{a}-\phi_{i-1}^{a}\phi_{i+1}^{b}\right)\right]\nonumber \\
 &  & +\frac{1}{2}\lambda_{6}\left[q_{i+1}^{a}(\phi_{i}^{a}\phi_{i+2}^{a}+\phi_{i}^{b}\phi_{i+2}^{b})-q_{i+1}^{b}\left(\phi_{i}^{b}\phi_{i+2}^{a}-\phi_{i}^{a}\phi_{i+2}^{b}\right)\right],\label{eq:derivVqa}\end{eqnarray}
 and \begin{eqnarray}
\frac{\partial V}{\partial q_{i}^{b}} & = & 2M^{2}q_{i}^{b}+2\lambda_{2}\left[(q_{i}^{a})^{2}+(q_{i}^{b})^{2}\right]q_{i}^{b}+2\lambda_{3}q_{i}^{b}\left[\sum_{j=1}^{N}(\phi_{j}^{a})^{2}+(\phi_{j}^{b})^{2}\right]\nonumber \\
 &  & +\mu(-\phi_{i-1}^{b}\phi_{i}^{a}+\phi_{i-1}^{a}\phi_{i}^{b})+2\lambda_{5}q_{i}^{b}\left[\sum_{j\neq i}(q_{j}^{a})^{2}+(q_{j}^{b})^{2}\right]\nonumber \\
 &  & +\frac{1}{2}\lambda_{6}\left[-q_{i-1}^{b}(\phi_{i-1}^{a}\phi_{i+1}^{a}+\phi_{i-1}^{b}\phi_{i+1}^{b})-q_{i-1}^{a}\left(\phi_{i-1}^{b}\phi_{i+1}^{a}-\phi_{i-1}^{a}\phi_{i+1}^{b}\right)\right]\nonumber \\
 &  & +\frac{1}{2}\lambda_{6}\left[-q_{i+1}^{b}(\phi_{i}^{a}\phi_{i+2}^{a}+\phi_{i}^{b}\phi_{i+2}^{b})-q_{i+1}^{a}\left(\phi_{i}^{b}\phi_{i+2}^{a}-\phi_{i}^{a}\phi_{i+2}^{b}\right)\right],\label{eq:derivVqb}\end{eqnarray}
 which leads for the VEVs in Eq.~(\ref{eq:universalVEVs}) to the
minimization condition \[
u\left[M^{2}+\left(\lambda_{2}+(N-1)\lambda_{5}\right)u^{2}+(N\lambda_{3}+\frac{1}{2}\lambda_{6})v^{2}\right]+\frac{1}{2}\mu v^{2}=0,\]
 and $\langle\partial V/\partial q_{i}^{b}\rangle=0$ is again satisfied
for these VEVs.

\section{Energy density and pressure of the quantized scalar field\label{secA:ScalarEngPress}}

In Sec.~\ref{sub:CasimirScalar}, we have introduced a real 5D bulk-scalar
$\phi$ propagating in a $\mathcal{M}\times\mathcal{S}_{\textrm{lat}}^{1}$
manifold. To normalize the field modes in Eq.~(\ref{eq:ScalarFieldAnsatz}),
we define the following scalar product for two modes $\phi_{1,2}(\vec{p},n)$
by\[
(\phi_{1},\phi_{2}):=i\int d^{3}x\, a\sum_{j=1}^{N}\,\left(\phi_{1}^{\dag}(\partial_{t}\phi_{2})-(\partial_{t}\phi_{1})^{\dag}\phi_{2}\right),\]
so that the normalization factor~$A$ can be fixed by demanding the
orthonormality relation\[
\left(\phi(\vec{p},n),\phi(\vec{p}^{\,\prime},n^{\prime})\right)=-V_{3}^{-1}\delta^{(3)}(\vec{p}-\vec{p}^{\,\prime})\delta_{nn^{\prime}},\]
where~$V_{3}$ is an arbitrary $3$-volume factor, which leaves the
scalar product dimensionless. With the Ansatz in Eq.~(\ref{eq:ScalarFieldAnsatz}),
we find\begin{equation}
A^{\dag}A=\frac{1}{2\omega(2\pi)^{3}V_{3}R},\label{eq:NormalizScalar}\end{equation}
where we have applied the relations\[
\int d^{3}x\cdot\exp(i\vec{x}(\vec{p}-\vec{p}^{\,\prime}))=(2\pi)^{3}\delta^{(3)}(\vec{p}-\vec{p}^{\,\prime})\,\,\,\,\textrm{and}\,\,\,\,\sum_{j=1}^{N}\exp\left(2\pi i\frac{n-n^{\prime}}{N}j\right)=N\delta_{nn^{\prime}}.\]
Now, the $00$-component $\hat{T}_{00}$ and the averaged $ii$-components
$\frac{1}{3}\sum_{i=1}^{3}\hat{T}_{ii}$ of the energy-momentum operator
$\hat{T}_{AB}$ follow from substituting the field operator $\hat{\phi}$
in Eq.~(\ref{eq:ScalarFeldop}) into Eqs.~(\ref{eq:EngDensClass})
and~(\ref{eq:PressClass}). Here, it is useful to consider the relations\begin{eqnarray}
(\partial_{0}\hat{\phi})^{\dag}(\partial_{0}\hat{\phi}) & = & V_{3}^{2}\int d^{3}p\int d^{3}p^{\prime}\sum_{n,n^{\prime}=1}^{N}\left[\partial_{0}\phi(\vec{p},n)a_{\vec{p},n}\cdot\partial_{0}\phi(\vec{p}^{\,\prime},n^{\prime})^{\dag}a_{\vec{p}^{\,\prime},n^{\prime}}^{\dag}+\cdots\right]\nonumber \\
 & = & V_{3}\int d^{3}p\sum_{n=1}^{N}\, A^{\dag}A\cdot\omega^{2}+\cdots\nonumber \\
(\partial_{b}\hat{\phi})^{\dag}(\partial_{b}\hat{\phi}) & = & V_{3}\int d^{3}p\sum_{n=1}^{N}\, A^{\dag}A\cdot p_{b}^{2}+\cdots\,\,\,\,\forall\, b=1,2,3\nonumber \\
(\partial_{5}\hat{\phi})^{\dag}(\partial_{5}\hat{\phi}) & = & V_{3}\int d^{3}p\sum_{n=1}^{N}\, A^{\dag}A\cdot2a^{-2}(1-\cos qa)+\cdots,\label{eq:ScalarTABParts}\end{eqnarray}
where the ellipses ($\cdots$) denote the terms which vanish in the
VEVs~$\left\langle 0\right|\hat{T}_{AB}\left|0\right\rangle $ due
to~$\left\langle 0\right|a^{\dag}=a\left|0\right\rangle =0$. When
we insert the terms in Eqs.~(\ref{eq:ScalarTABParts}) into Eqs.~(\ref{eq:EngDensClass})
and~(\ref{eq:PressClass}), one obtains, after taking the VEVs of
$\hat{T}_{00}$ and $\hat{T}_{ii}$, the energy density~$\rho_{5}$
and the pressure~$p_{5}$ of the quantized field $\phi$,\begin{eqnarray*}
\rho_{5}=\left\langle 0\right|\hat{T}_{00}\left|0\right\rangle  & = & V_{3}\int\textrm{d}^{3}p\sum_{n=1}^{N}\, A^{\dag}A\cdot\left[\frac{1}{2}\omega^{2}+\frac{1}{2}\vec{p}^{\,2}+\frac{1}{2}\cdot2a^{-2}(1-\cos qa)+\frac{1}{2}M_{\textrm{s}}^{\textrm{2}}\right]\\
 & = & V_{3}\int\textrm{d}^{3}p\sum_{n=1}^{N}\, A^{\dag}A\cdot\omega^{2},\end{eqnarray*}
\begin{eqnarray*}
p_{5}=\frac{1}{3}\sum_{i=1}^{3}\left\langle 0\right|\hat{T}_{ii}\left|0\right\rangle  & = & V_{3}\int\textrm{d}^{3}p\sum_{n=1}^{N}\, A^{\dag}A\cdot\left[\frac{1}{3}\vec{p}^{\,2}+\frac{1}{2}\omega^{2}-\frac{1}{2}\vec{p}^{\,2}-\frac{1}{2}2a^{-2}(1-\cos qa)-\frac{1}{2}M_{\textrm{s}}^{2}\right]\\
 & = & V_{3}\int\textrm{d}^{3}p\sum_{n=1}^{N}\, A^{\dag}A\cdot\frac{\vec{p}^{\,2}}{3\omega},\end{eqnarray*}
where we have used the energy-momentum relation $\omega^{2}=\vec{p}^{\,2}+2a^{-2}(1-\cos qa)+M_{\textrm{s}}^{2}$.
With the normalization factor~$A$ in Eq.~(\ref{eq:NormalizScalar})
we finally arrive at the Eqs.~(\ref{eq:ScalarRhoUnregul}) and~(\ref{eq:ScalarPUnregul}).

\section{Renormalization\label{secA:Renormalization}}

To gain the finite and unambiguous Casimir energy density, it is necessary
to compare the discrete mode sums belonging to the momenta in~$\mathcal{S}_{\textrm{lat}}^{1}$
with the energy density and pressure of a field in a space-time with
a non-compactified extra dimension. Regarding Eqs.~(\ref{eq:ScalarRhoUnregul})
and~(\ref{eq:ScalarRhoRegul}), the mode sum with respect to the
5th momentum coordinate~$q=2\pi n/(aN)$ is of the type\[
\sum_{n=1}^{N}f(n/N),\]
where~$f(n/N)$ summarizes the terms in the last line of Eq.~(\ref{eq:ScalarRhoRegul}).
From this sum, the mode integral corresponding to a non-compactified~$\mathbb{R}^{1}$-dimension
can be obtained by cutting out a section of length~$R$ of an~$\mathbb{R}^{1}$-dimension.
This means, that we take the limit of an infinite number~$M$ of
lattice sites,~$M\rightarrow\infty$, while keeping the spacing~$a$
constant: \[
\left.\frac{R}{Ma}\sum_{n=1}^{M}f\left(\frac{n}{M}\right)\right|_{M\rightarrow\infty}=\frac{R}{a}\left[\sum_{n=1}^{M}\frac{\Delta n}{M}f\left(\frac{n}{M}\right)\right]_{M\rightarrow\infty}\stackrel{s:=n/M}{=}N\cdot\int_{0}^{1}\textrm{d}s\cdot f(s),\]
where~$Ma$ becomes the infinite {}``length'' of~$\mathbb{R}^{1}$
and~$f(s)$ is the same function as in the~$\mathcal{S}_{\textrm{lat}}^{1}$
mode sum. In the last equation, we have substituted~$s:=n/M$ and
inserted~$\Delta n=1$ so that~$\textrm{d}s=\Delta n/M$ for~$M\rightarrow\infty$.
Both the sum and the integral are finite since the lattice introduces
an UV cutoff. Then the renormalization is performed by subtracting
the integral from the sum, \begin{equation}
\sum_{n=1}^{N}f\left(\frac{n}{N}\right)-N\cdot\int_{0}^{1}\textrm{d}s\cdot f(s)=\sum_{n=1}^{N}m^{4}\ln m-N\cdot\int_{0}^{1}\textrm{d}s\cdot m^{4}\ln m,\label{eq:Sum1NMinusInt01}\end{equation}
where only~$m^{4}\ln m$ survives since all other terms\[
\frac{1}{2}d^{-4}+\frac{1}{4}m^{2}d^{-2}+\frac{1}{8}m^{4}\left(\frac{1}{4}+\frac{1}{2}\gamma-\ln2+\ln(d)\right)+\mathcal{O}(d^{6}m^{6})\]
of Eq.~(\ref{eq:ScalarRhoRegul}) either vanish when the regularization
is removed for $d\rightarrow0$ or are completely subtracted due to
the following identities:\begin{equation}
\sum_{n=1}^{N}(1-\cos2\pi\kl{\frac{n}{N}}+\kl{\frac{1}{2}}a^{2}M_{\textrm{s}}^{2})=N\cdot\int_{0}^{1}ds\cdot(1-\cos2\pi s+\kl{\frac{1}{2}}a^{2}M_{\textrm{s}}^{2})=N(1+\kl{\frac{1}{2}}a^{2}M_{\textrm{s}}^{2}),\label{eq:RenTerm2}\end{equation}
 \begin{equation}
\sum_{n=1}^{N}(1-\cos2\pi\kl{\frac{n}{N}}+\kl{\frac{1}{2}}a^{2}M_{\textrm{s}}^{2})^{2}=N\cdot\int_{0}^{1}ds\cdot(1-\cos2\pi s+\kl{\frac{1}{2}}a^{2}M_{\textrm{s}}^{2})^{2}=N(\kl{\frac{3}{2}}+a^{2}M_{\textrm{s}}^{2}+(\kl{\frac{1}{2}}a^{2}M_{\textrm{s}}^{2})^{2}).\label{eq:RenTerm3}\end{equation}
 This is also the case for twisted fields, where~$n$ is replaced
by~$n-\frac{1}{2}$.

\end{document}